\documentclass[11pt, a4paper]{article}
\usepackage{subcaption}
\usepackage{graphicx}
\usepackage{amssymb}
\usepackage{longtable}
\usepackage{booktabs}
\usepackage[colorlinks=true, breaklinks=true]{hyperref}
\AtBeginDocument{%
\hypersetup{
    linkcolor=blue,
    filecolor=magenta,      
    urlcolor=cyan,
    pdftitle={Test case generation for agent-based models: A systematic literature review},
    pdfpagemode=FullScreen,
    citecolor=black
}}
\usepackage{xcolor}
\usepackage{listings}
\usepackage[modulo]{lineno}
\usepackage[export]{adjustbox}
\usepackage{caption}
\usepackage{macros/codebox}
\usepackage{newunicodechar}
\usepackage{amsmath}
\usepackage{xltabular}
\usepackage[titletoc]{appendix}
\usepackage{todo}
\newcommand{\doinote}[1]{{%
  \let\thempfn\relax
  \footnotetext[0]{#1}
}}

\author{\small
  Andrew G. Clark,
  Neil Walkinshaw,
  Robert M. Hierons\\ \small
  The University of Sheffield, UK\\
  \small
  \texttt{\{agclark2, n.walkinshaw, r.hierons\}@sheffield.ac.uk}
}
\title{\textbf{Test case generation for agent-based models: A systematic literature review}}
\date{}

\begin{document}

\maketitle
\doinote{DOI: \url{10.1016/j.infsof.2021.106567}}
\begin{abstract}
\footnotesize
\noindent\textbf{Context:} Agent-based models play an important role in simulating complex emergent phenomena and supporting critical decisions. In this context, a software fault may result in poorly informed decisions that lead to disastrous consequences. The ability to rigorously test these models is therefore essential.\newline

\noindent\textbf{Objective:} Our objective is to summarise the state-of-the-art techniques for test case generation in agent-based models and identify future research directions.\newline

\noindent\textbf{Method:} We have conducted a systematic literature review in which we pose five research questions related to the key aspects of test case generation in agent-based models: What are the information artifacts used to generate tests? How are these tests generated? How is a verdict assigned to a generated test? How is the adequacy of a generated test suite measured? What level of abstraction of an agent-based model is targeted by a generated test?\newline

\noindent\textbf{Results:} Out of the 464 initial search results, we identified 24 primary publications. Based on these primary publications, we formed a taxonomy to summarise the state-of-the-art techniques for test case generation in agent-based models. Our results show that whilst the majority of techniques are effective for testing functional requirements at the agent and integration levels of abstraction, there are comparatively few techniques capable of testing society-level behaviour. Furthermore, the majority of techniques cannot test non-functional requirements or ``soft goals''.\newline

\noindent\textbf{Conclusions:} This paper reports insights into the key developments and open challenges concerning test case generation in agent-based models that may be of interest to both researchers and practitioners. In particular, we identify the need for test case generation techniques that focus on societal and non-functional behaviour, and a more thorough evaluation using realistic case studies that feature challenging properties associated with a typical agent-based model.
\end{abstract}

\section{Introduction}
In recent years, computational models have become increasingly popular in a variety of domains, with applications ranging from economic modelling \cite{liu2020interbank} to public health \cite{tracy2018agent}. The ability to model and understand vast quantities of data, coupled with the increased availability of data, have helped to establish the computational model as an essential research tool. Agent-based models are one variety of computational model that are used to simulate complex scenarios involving many individuals, often to inform social and economic policies, including national responses to the recent COVID-19 pandemic \cite{flaxman2020report, panovska2020determining, kerr2020covasim}. With such critical applications, poor modelling assumptions or software faults can lead to serious repercussions.
Software testing is an essential software engineering practice that aims to ensure the quality of software products, including the identification and removal of faults. The time and effort necessary to test software and achieve such quality are significant, often accounting for over 50\% of total project costs \cite{ramler2006economic}. In recent years, a variety of techniques have been proposed to alleviate the cost of testing, from search-based testing \cite{mcminn2004search} to fuzzing \cite{godefroid2008automated}, that automate or partially-automate various testing activities such as the creation of test cases.

Of all software testing activities, perhaps the most labour-intensive and time-consuming activity is the creation of test cases \cite{prasanna2005survey}. As a consequence, the topic of test case generation has received significant attention in recent years, leading to a variety of surveys, techniques and tools supporting the automatic or semi-automatic derivation of test suites \cite{anand2013orchestrated}. However, for many real-world applications, it remains unclear how effective these tools and techniques are, with areas such as scientific software requiring further empirical study to establish their applicability \cite{kanewala2014testing}.

Agent-based modelling is one area where the applicability of existing test case generation techniques is unclear. Due to their often non-deterministic nature coupled with their complex emergent behaviour, agent-based models present significant challenges for test case generation \cite{russell2002artificial, luck2005agent}. For example, in response to the ongoing COVID-19 pandemic, the UK government has used an agent-based model developed at Imperial College London to inform an economic policy estimated to cost \pounds192.3bn over the 2020-2021 financial year \cite{flaxman2020report, obr2020coronavirus:online}. Despite the importance of testing such critical software, the model contains only basic integration tests configured to a specific set of parameters \cite{mrcideco79:online}. To this end, there have been several recent calls for improved techniques for validation and testing of predictive models \cite{squazzoni2020computational, wynants2020prediction}.

Clearly, there is a need for tools and techniques that make software testing an accessible practice for both researchers and practitioners regardless of their discipline. In this review we aim to identify the current state-of-the-art techniques and future research directions for test case generation in agent-based modelling, paying particular attention to their potential real-world applications. In order to achieve a fair and comprehensive survey of the existing test case generation literature in the agent-based modelling field, this paper reports the results of a systematic literature review in accordance to the guidelines proposed by Kitchenham and Charters \cite{kitchenham2007guidelines}.

The remainder of this work is organised as follows: in Section 2, we provide an overview of agent-based models and test case generation. In Section 3, we present the research methodology including the research questions and review procedure. In Section 4, the results of the systematic literature review are presented, followed by a discussion of some particularly interesting observations in Section 5. In Section 6 we conclude the review and present potential future directions for research.

\section{Background}
In this section, we introduce the context and core concepts that concern this review, namely test case generation and agent-based models. An informal definition of test case generation will be provided and then revisited in the context of an agent-based model. In addition, the challenges associated with testing an agent-based model and their impact on the test case generation process are highlighted.

\subsection{Agent-based models}
An agent-based model is a system of autonomous processes characterised by sets of individual and collective goals which, upon execution, typically result in complex emergent behaviour. Whilst there is no universal definition of an agent, there are several criteria introduced by Crooks et al. \cite{crooks2012introduction} that are generally agreed upon as defining features of an agent:

\begin{enumerate}
  \item Autonomy: agents are self-governing individuals that make decisions and act without centralised influence.
  \item Heterogeneity: agents are diverse individuals that are distinguished by differences in their characteristics (e.g. age, gender, height).
  \item Active: agents percept and affect their environment, including other agents, to achieve their goals.
\end{enumerate}

Many different forms of agent-based model and similar paradigms exist, from the beliefs, desires and intentions framework \cite{rao1995bdi} (BDI agents) to cellular automata \cite{wolfram1983statistical},  however, in practice the concepts tend to overlap. In this review we consider an agent-based model to be an umbrella term that captures all of the above criteria.\footnote{For reasons of practicality a line is drawn at physical and cyber-physical systems as these topics fall into the field of robotics, though one could argue that they also fall into the agent-based category.}

Agent-based models lend themselves to modelling theory-driven phenomena \cite{arnold2019dag} such as the impact of social policies \cite{chao2015dynamic} or the spread of infectious disease \cite{frias2011agent}, where the underpinning mechanisms are well-understood, but their larger-scale implications are not. A recent example of a significant agent-based model is the COVID-19 pandemic model developed at Imperial College London; a complex simulation that has been used to inform the United Kingdom's response to the COVID-19 pandemic \cite{flaxman2020report}. However, COVID-19 models have received criticism \cite{chatterjee2020transparency, wynants2020prediction} highlighting a lack of transparency, reproducibility, and an urgent need for more rigorous validation. To this end, software testing techniques may be applied to generate test cases that exercise the system under different conditions to reveal software faults and validate the model.

\subsection{Test case generation}
Informally, test case generation is the activity in which test cases are created either manually, semi-automatically or automatically to expose faulty behaviour in a software system. In general, an approach to testing relies on four key components: an information artifact, a generation mechanism, test data adequacy criteria and a test oracle. Here we define these components alongside fundamental testing concepts that will be mentioned throughout the paper and present an overview of the test case generation process used to frame and motivate our research questions.\newline

\begin{itemize}
  \item \textbf{Test} - The act of exercising software with test cases with the goal of discovering faults or demonstrating correctness \cite{jorgensen2018software}.
  \item \textbf{Test case} - A set of inputs, preconditions and expected outcomes associated with a particular program behaviour \cite{jorgensen2018software}.
  \item \textbf{Test suite} - A collection of test cases \cite{harrold2001regression}.
  \item \textbf{Test execution} - An execution of the system-under-test constrained to the conditions specified by a given test case.
  \item \textbf{Information artifact} - A source of information from which test cases may be derived. An information artifact can take the form of a formal specification, a design document or source code, and provides information about the functionality or goals of the system that are necessary to produce meaningful test cases.
  \item \textbf{Generation mechanism} - An algorithm or strategy used to produce a test case from an information artifact. A generation mechanism is central to the creation of test cases, such as the application of mutation operators to seed test cases or the use of user-defined input generation rules.
  \item \textbf{Test data adequacy criterion} - A measure of test case quality with respect to a testing goal such as coverage \cite{zhu1997software}. A test data adequacy criterion is often used to guide test case generation towards achieving the defined goal.
  \item \textbf{Test oracle} - A procedure that determines whether a given test execution of the system-under-test is correct or not \cite{mcminn2015oracle}. A test oracle may be specified, derived, implicit, or human.
\end{itemize}

In light of these definitions, test case generation may be described as the process by which a test suite is generated from an information artifact using a generation mechanism, often guided by a test data adequacy criterion. Following the generation of a test case, its execution on the system-under-test should be judged by a test oracle to determine whether it passes or fails. However, it is important to note that these features are not essential for test case generation, but are necessary to consider as part of the larger testing process. A generalised overview of the testing process is presented in Figure \ref{fig:tcgProcess} below.

\begin{figure}[!htb]
  \centering
  \includegraphics[width=\linewidth]{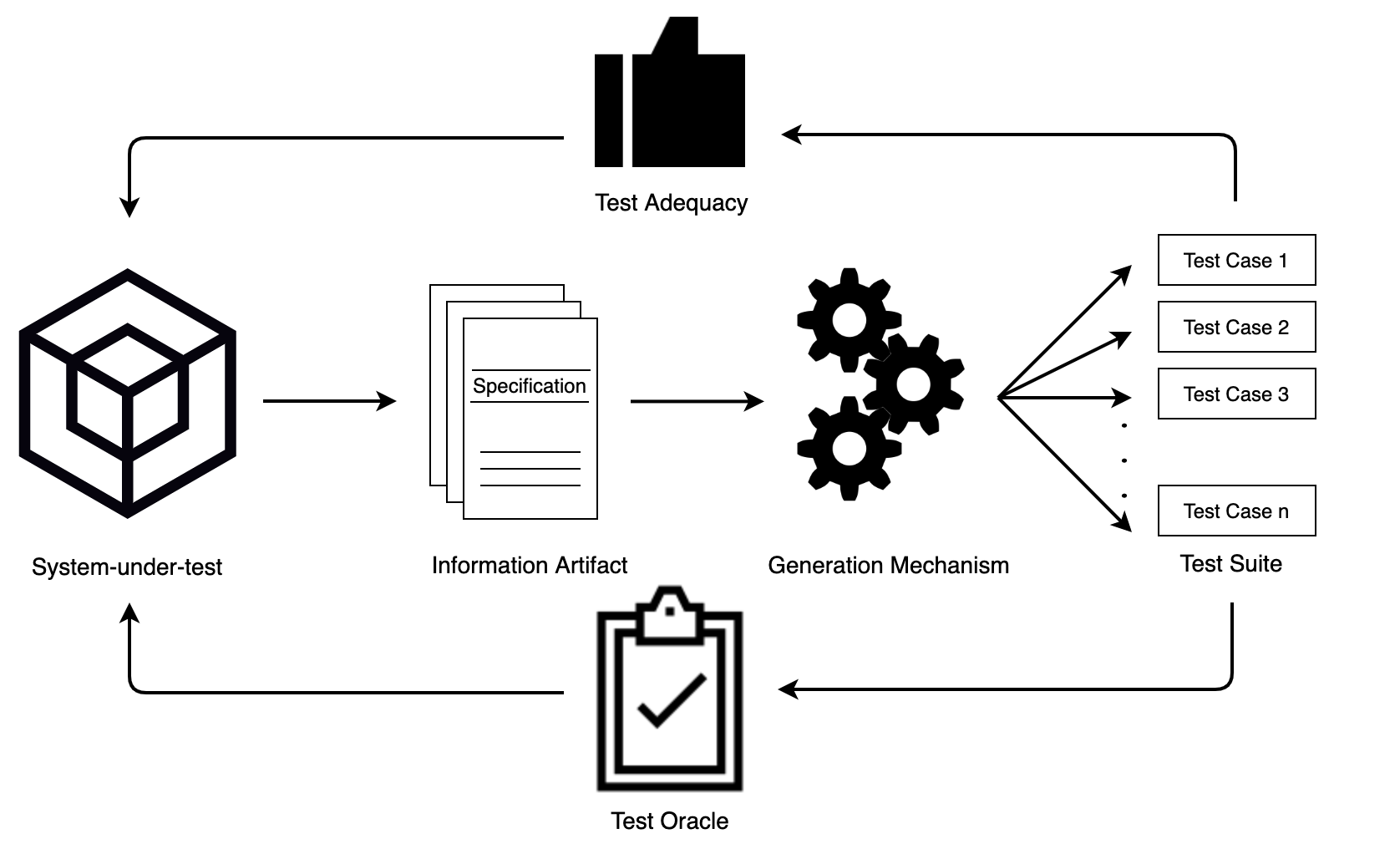}
  \caption{An overview of the testing process.}
  \label{fig:tcgProcess}
\end{figure}

\subsection{Testing agent-based models}
Software testing is a significant challenge in the context of agent-based models. This is partly due to the challenging features of an agent-based model \cite{nguyen2008experimental}, but may also be attributed to the domain in which agent-based models are typically developed, such as epidemiology and social sciences, where there are few formally trained software engineers with testing experience \cite{kanewala2014testing}. Additionally, with multiple levels of abstraction, an agent-based model requires the consideration of several different testing objectives. Nguyen et al. propose five levels of abstraction to consider when testing agent-based models \cite{nguyen2009testing}, these are as follows:

\begin{enumerate}
  \item Unit: Testing of individual units that make up an agent, including code, goals and plans.
  \item Agent: Testing the integration of the different units of an agent, amounting to their ability to act as an individual.
  \item Integration: Testing the interaction of agents, the interaction of agents with their environment, and communication protocol and semantics.
  \item Society: Testing the multi-agent system as a collective, including the macroscopic and emergent behaviour of the overall system.
  \item Acceptance: Testing the multi-agent system in the real execution environment to verify whether the implementation meets stakeholder goals.
\end{enumerate}

Agent-based models have unique properties that pose significant challenges for software testing. These challenges affect all levels of abstraction, including individual agents, interacting agents, and the resulting society of agents.

Agents are autonomous; they can proactively plan and adapt their behaviour to achieve specific goals \cite{padgham2005developing}. With control over their state \cite{adra2010mutation}, agents may derive new plans of their own accord to tackle problems in previously unknown ways \cite{tiryaki2006sunit}. However, autonomous behaviour also exacerbates the test oracle problem \cite{mcminn2015oracle}: how can we determine whether newly introduced behaviour is correct or incorrect, given that it is absent in the specification?

Agents can interact with many other agents and their environment concurrently, leading to a potentially intractable number of different interactions and scenarios \cite{zheng2005conformance}. As a consequence, integration-level testing suffers from the problem of controllability \cite{freedman1991testability}. Due to the large number of autonomous interactions, it is difficult to instantiate test cases that will consistently lead to the desired output. Therefore, it is particularly difficult to control and test social behaviour in an agent-based system.

On a macroscopic scale, autonomy and agent interaction lead to complex patterns of social behaviour, known as emergent behaviour. Emergent behaviour is difficult to test as it may require significant time to develop, is not always guaranteed to occur, and cannot be predicted from the characteristics of the involved agents alone \cite{scholl2001agent}. Additionally, a tester must first address the challenges that hinder the lower levels of testing to achieve society level testing.

Testing an agent-based model generally involves executing the model under a particular set of conditions for the agents and their environment. To ascertain whether the model is correct, the resulting outcome is then compared to the expected outcome, which may be captured in either a formal or informal specification. However, due to the wide variety of different agent-based modelling paradigms, frameworks, and tools, the testing process will vary. For example, a technique developed for testing the conformance of messages sent between message-passing agents may not be compatible with a model involving proximity-based interactions.

In the context of an agent-based model, test case generation involves the creation of valid and invalid input data that trigger a particular sequence of events \cite{nguyen2008ontology}. Test case generation may be manual, or may be automated in various ways to support the testing process outlined in Section 2.2. A comprehensive technique may also cover other aspects of the testing process, such as an agent-specific test adequacy criterion to guide test case generation towards achieving a testing goal. A test oracle may also be provided to assign verdicts to test cases.

\subsection{Related work}
To the best of the authors' knowledge, there are no existing systematic literature reviews on the topic of test case generation for agent-based models. However, there are several systematic reviews of related topics.

Bakar and Selamat \cite{bakar2018agent} conducted a systematic literature review and mapping on the topic of agent systems verification, focusing on techniques for checking agent properties and fault detection during the stages of the agent-based development process. In comparison to our own literature review, Bakar and Selamat's review has a wider scope as it investigates testing techniques for agent systems in general, mapping the different properties and faults that existing techniques may detect, whereas our review focuses on test case generation specifically and explores this process in greater detail. Despite having a different scope, many of the results from Bakar and Selmat's review support our own observations. For example, Bakar and Selmat's mapping of the types of properties tested by existing techniques reveal that functional requirements can be tested using many techniques, whilst non-functional requirements cannot. This aligns with our own observation that techniques for generating functional test cases are well-established, whilst there are fewer techniques that target non-functional requirements.

Blanes et al. \cite{blanes2009requirements} present a systematic review of requirements engineering in the development of multi-agent systems. Whilst this work does not focus on testing or test case generation for agent-based models, it covers different aspects of requirements engineering including specification and validation that have some relevance to our own review. In this work, ontologies are recognised as a technique to deal with requirements, supporting the use of ontologies as a formal information artifact in our taxonomy. Blanes et al. also indentified that roughly a quarter of the literature they reviewed provided some form of automated validation, highlighting the need for new automatic approaches to test agent-based models.

Arora and Bhatia \cite{arora2018systematic} conduct a systematic review of agent-based test case generation for regression testing. Whilst this approach is relevant to test case generation, it reviews agent-based approaches to test case generation, whereas our literature review is concerned with test case generation approaches for agent-based models. That is, this review investigates the use of agents for testing rather than how agents can be tested. Although this review does not have any significant relevance to our own, it is necessary to highlight the difference as, based on the title alone, it would appear to be similar.

Existing reviews have covered verification in agent systems, requirements engineering for multi-agent systems, and the use of agents to generate test cases for regression testing. However, these reviews do not consider the components supporting test case generation, such as specification languages or test oracles, in-depth. Our review addresses this gap by reviewing the artifacts used to form tests cases, generation mechanisms, test oracles, test adequacy metrics, and levels of testing abstraction achieved. To that end, our review covers not only the state-of-the-art test case generation techniques for agent-based models but also the individual components on which these techniques depend.
\section{Research methodology}
In recent years agent-based technology has increased in popularity across a variety of domains ranging from epidemiology \cite{keyes2017invited} to economics \cite{platt2020comparison}. Meanwhile, significant advances in software testing have revolutionised the way we test complex systems, such as the emergence of fuzzing techniques to enhance the automatic discovery of security vulnerabilities \cite{godefroid2012sage}. Despite significant progress in both agent-based modelling and software testing, the need for better validation of agent-based models has been highlighted recently \cite{fagiolo2019validation, squazzoni2020computational, wynants2020prediction}.

This paper reviews existing state-of-the-art test case generation techniques for agent-based models. In this section, we outline the research methodology used to conduct the review, including the review protocol and our research questions.

\subsection{Research questions}
We pose five research questions that focus on the individual aspects of the testing process outlined in Section 2.2. To help answer each question, a taxonomy grouping similar techniques will be introduced in Section 4, facilitating a collective discussion of the key similarities and differences amongst the state-of-the-art.

\subsubsection{RQ1: What artifacts are used to drive test case generation for agent-based models?}
An approach to test case generation requires knowledge of the system-under-test if it is to produce meaningful inputs, identify errors or measure the adequacy of the generated test cases. This knowledge is provided by some information artifact (defined in Section 2.2). In this research question, we identify the information artifacts specific to agent-based models that have been used to drive test case generation.

\subsubsection{RQ2: What mechanisms are used to generate tests?}
An approach to test case generation requires some form of generation mechanism (defined in Section 2.2) that can utilise the knowledge contained in an information artifact to create test cases.  Given that there are different types of agent-based models, each with potentially different purposes, it follows that there are also many different forms of generation mechanisms. Additionally, due to the lack of software testing experience amongst scientific developers \cite{kanewala2014testing}, commonplace approaches to test case generation that rely on knowledge of specific technology or languages, such as unit testing in JUnit, may be unsuitable for testing agent-based models. In this research question, we look at the existing approaches used to generate test cases for agent-based models.

\subsubsection{RQ3: How are verdicts assigned to generated test cases?}
Given the often exploratory nature of agent-based models \cite{thaler2019show}, the expected outcome of an execution may be unknown and it may be difficult to assign a verdict to a test case. This is known as the test oracle problem \cite{mcminn2015oracle}. In evaluating an approach to test case generation it is essential to consider the test oracle, as a test case without a verdict provides no information about the validity or quality of the system-under-test. This issue has been discussed in detail by Staats et al. \cite{staats2011programs}. In the context of agent-based models, it is unclear how oracles are used to assign verdicts to test cases. In this research question, we investigate this issue.

\subsubsection{RQ4: How is the adequacy of a generated test suite measured?}
Test data adequacy is a measure of how good a test suite is with respect to some testing goal, such as coverage (see Section 2.2). The definition of a test data adequacy criterion is essential to the fault detection capabilities of a generated test suite \cite{zhu1997software}, as it can be used to drive test case generation towards areas of the system-under-test that are yet to be tested. It follows from our previous research questions that different test case generation techniques will involve different information artifacts and mechanisms, and therefore are likely to have different testing goals too. In this research question, we aim to identify the test data adequacy criterion used to judge the extent to which such goals have been met.

\subsubsection{RQ5: What level of abstraction do the generated test cases target?}
Due to the complex, multi-layered nature of agent-based models, Nguyen et al. propose five levels of abstraction in terms of which agent-based models should be tested: unit, agent, integration, society and acceptance as described in Section 2.3. Previous work has identified a lack of testing approaches for agent-based models that target the society and acceptance levels of testing \cite{nguyen2009testing}. However, this is yet to be investigated for the activity of test case generation. In this research question, we aim to investigate the different levels of abstraction that generated test suites are capable of testing.

\subsection{Review protocol}
To conduct this systematic literature review, we followed the guidelines presented by Kitchenham et al. \cite{kitchenham2007guidelines} and Petersen et al. \cite{Petersen2015}. This involved the following steps:

\begin{itemize}
  \item Selecting data sources from which the primary publications were collected.
  \item Constructing a search string to capture all relevant publications.
  \item Defining inclusion and exclusion criteria to systematically determine the relevance of publications.
  \item Designing a quality assessment questionnaire to assess the quality of the relevant publications with respect to the goals of the review.
\end{itemize}

\begin{figure}[h]
  \centering
  \includegraphics[width=\linewidth]{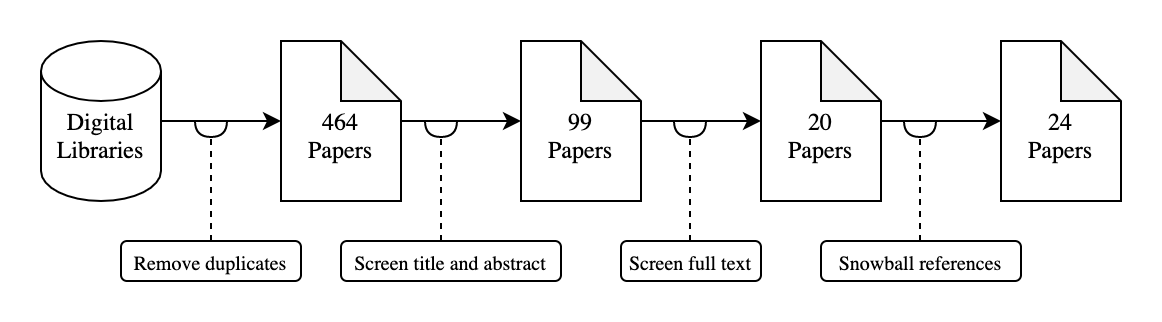}
  \caption{Overview of the search procedure.}
  \label{fig:searchProcess}
\end{figure}

To conduct the search procedure, we initially considered the title and abstract of all papers retrieved using the search string, removing only those that were irrelevant according to the screening criteria. In the next phase we considered the full-text of the remaining papers, and where there was any doubt over the relevance of a study, each researcher would independently review and discuss the paper until a consensus could be met. After identifying the primary publications, forward snowballing was conducted on the primary publications by searching their references to identify any literature missed by the search that may satisfy the screening criteria. The search procedure was then repeated for the snowballed papers. To reduce any potential bias, the components of the protocol have been independently reviewed by each author and a validation procedure has been conducted where necessary. An overview of the search procedure is given in Figure \ref{fig:searchProcess}.

In the following sections, we will explain each stage of the review protocol in greater detail.

\subsection{Data sources}
Following the guidelines presented by Petersen et al. \cite{Petersen2015}, we selected five reputable digital libraries as our data sources, including both scientific and indexed databases. The data sources were selected based on their coverage of high-impact agent-based test case generation papers, in addition to the search functionality offered. To validate the choice of data sources, each author identified a number of papers that were relevant to the goals of the review, and we ensured that each paper was available from at least one of the data sources. The selected data sources alongside the number of studies retrieved from each are shown in Table \ref{tab:retrievedStudies}.

\begin{table}[htbp]
\centering
\begin{tabular}{lll}
\toprule
Digital Library & URL & Studies Retrieved  \\ \midrule
ScienceDirect & \url{https://www.sciencedirect.com/} & 77 \\
IEEE & \url{https://ieeexplore.ieee.org/} & 58 \\
Scopus & \url{https://www.scopus.com/} & 42  \\
ACM & \url{https://dl.acm.org/} & 91  \\
SpringerLink & \url{https://link.springer.com/} & 196 \\ \midrule
Total & &  464  \\
\bottomrule
\end{tabular}
\caption{Literature retrieved from each data source.}
\label{tab:retrievedStudies}
\end{table}

\subsection{Search string}
We defined a targeted search string to retrieve the relevant studies from each data source. To form the search string, we adapted the \emph{PICO} criteria suggested by Kitchenham et al. \cite{kitchenham2007guidelines}. \emph{PICO} is an acronym designed to help produce a more effective search string by breaking up the focus of the search into the categories: \emph{Population}, \emph{Intervention}, \emph{Comparison} and \emph{Outcome}.

\emph{Population} concerns the target group of the publication, such as a certain demographic, industry or users of a particular tool. For this survey, we identified the population as publications that involve a form of agent-based model.

\emph{Intervention} covers the technique, technology or software used to solve a particular problem in the context of the selected population. We identified the intervention as a test case generation technique, given that we are investigating test case generation for agent-based models.

\emph{Comparison} concerns an alternative intervention that is to be compared with the identified intervention. Given that this review is exploratory in nature, we decided that it would not be appropriate to include the comparison category in the search. For instance, if we included \emph{``search-based''} and \emph{``random testing''} as terms for comparison, any literature that does not mention one of these techniques would be excluded.

\emph{Outcome} relates to the qualities or results of the study focus that a researcher or practitioner would identify as important. In this study, we identified the outcome as a test, given that any test case generation technique should produce at least one test case if successful.

Using the above criteria, we formulated the following search string by identifying synonyms for each group of search terms, joining these by the boolean $OR$ operator, and then joining the resulting clauses with the boolean $AND$ operator, as shown below in Listing \ref{lst:searchString}. The one exception is the population search terms, which are split into two $OR$ clauses joined by an $AND$, such that a study may refer to the population as an agent-based model, agent-based system, agent-based simulation, multi-agent system, multi-agent model, or a multi-agent simulation.


\begin{figure}
  \begin{codebox}{Java}{Search string for retrieving relevant studies}{lst:searchString}
    ("agent-based" OR "multi-agent") AND
    ("system" OR "model" OR "simulation") AND
    ("test generation" OR "test case generation" OR "test input generation" OR "parameter selection") AND "test"
  \end{codebox}
\end{figure}

Similar to the validation procedure employed for the selection of data sources, each researcher independently identified a collection of papers relevant to the literature review that was used as a benchmark to judge the validity of the search string. The search string was modified until it was capable of obtaining all of the benchmark studies from the selected data sources, resulting in the search string displayed in Listing \ref{lst:searchString}. This included the separation of the population terms by the $OR$ operator as described above, inclusion of the \emph{``simulation''} term, and inclusion of the term \emph{``parameter selection''} which was often used to describe a process that resembles test case generation. During this process the format of the search string was also modified to fit the syntax of each individual search engine, preserving the same meaning as much as possible, within the constraints of the differing functionality of the data sources.

Whilst the validation procedure improves our confidence in the search string, it does not guarantee that it will capture all relevant papers. In fact, we identified a further 4 relevant papers through forward snowballing. These papers were not captured by the initial search as they included population or intervention terms that were too general or niche to include in the search string. Inclusion of the former would increase the rate of false positives, making it infeasible to review each paper. Inclusion of the latter would not have improved the precision beyond the snowballed paper, making the search string overly-specific (an issue for digital libraries constrained to a maximum number of search terms such as ScienceDirect).

\subsection{Screening criteria}
In order to systematically decide which of the retrieved studies were relevant, we applied the inclusion and exclusion criteria shown in Table \ref{tab:inclusionCriteria}. In order to satisfy the inclusion criteria, a paper must satisfy both \textbf{IC1} and \textbf{IC2} in addition to at least one of \textbf{IC3}, \textbf{IC4} or \textbf{IC5}. This ensures that a relevant paper focuses on the topic of test case generation for agent-based models and has been peer-reviewed. On the other hand, if any of the exclusion criteria are satisfied then the paper is deemed irrelevant. The criteria were applied during each phase of the search process, namely the title and abstract review, the full-text review and snowballing as shown in Figure \ref{fig:searchProcess}.

After applying the inclusion and exclusion criteria to an initial selection of 464 papers, we identified 20 relevant papers. A large number of false positives were obtained as there exists a significant body of research on the application of agents to testing activities, such as agent-based test case generation \cite{arora2018systematic}. This research concerns a similar but irrelevant topic and thus uses terminology that is difficult to distinguish from relevant research. To address this issue, we introduced a specific exclusion criterion: \textbf{EC2}.


\begin{table}[htbp]
\centering
\begin{tabular}{ll}
\toprule
ID & Criteria  \\ \midrule
\textbf{IC1} & Evaluates or proposes at least one test case generation \\ &  method for an agent-based model \\
\textbf{IC2} & Study is published in either a journal or conference \\ & proceedings \\
IC3 & Mentions how the expected outcomes for generated test \\ & cases are known i.e. whether a test case should pass \\ & or fail for a given input  \\
IC4 & Mentions the information artifact from which test cases \\ & are derived e.g. a specification \\
IC5 & Mentions a test adequacy measure \\
\midrule
EC1 & Focus of the study does not relate to test case generation \\ & for agent-based models \\
EC2 & Study involves using agent-based models to generate test \\ & cases for a system that is not agent-based \\
EC3 & Non-English study \\
EC4 & Duplicate study \\
EC5 & Study is unavailable in hard or electronic format \\
\bottomrule
\end{tabular}
\caption{Inclusion and exclusion criteria for determining relevance of retrieved papers.}
\label{tab:inclusionCriteria}
\end{table}


\subsection{Quality assessment}
To appraise the quality of the selected studies, we produced a quality assessment questionnaire adapted from the example questionnaires presented by Kitchenham and Charters \cite{kitchenham2007guidelines}, as shown in Table \ref{tab:qualityAssessmentCriteria}. We primarily judge the credibility of a study based on the clarity of its objectives and evaluation, its reproducibility and its real-world impact. Concerning each criterion, a study may be scored as either 0, 0.5 or 1, corresponding to not fulfilled, partially fulfilled and fulfilled respectively. To ensure the credibility of selected studies, we set a minimum quality threshold of 1.5 such that any paper with a quality score less than or equal to 1.5 is ignored due to a perceived lack of credibility.

\begin{table}[!htb]
\centering
\begin{tabular}{ll}
\toprule
ID & Criteria  \\ \midrule
QA1 & Are the aims of the study clear? \\
QA2 & Have all aims of the study been addressed in the \\ & evaluation? \\
QA3 & Is sufficient information available to reproduce the \\ & study? \\
QA4 & Does the study consider real-life applications? \\
\bottomrule
\end{tabular}
\caption{Quality assessment criteria for assessing the quality of relevant literature.}
\label{tab:qualityAssessmentCriteria}
\end{table}

%

\section{Results}
In this section, we outline the results for each research question according to a derived taxonomy. It is important to note that the purpose of the proposed taxonomy is to group and discuss similar approaches to test case generation for agent-based models, rather than to act as a concrete definition. As a result, there may be some overlap between categories within a taxonomy. A separate bibliography for the reviewed literature is included as an Appendix, where citations to this bibliography are prefixed with ``P'' e.g. [P1].

\subsection{RQ1: What artifacts are used to drive test case generation?}
All of the reviewed approaches use some form of specification as an information artifact to drive test case generation. The types of information artifact can be divided into two different categories: formal and informal artifacts.
\begin{figure}[h!]
\centering
\begin{subfigure}{0.9\textwidth} 
   \includegraphics[width=1\linewidth]{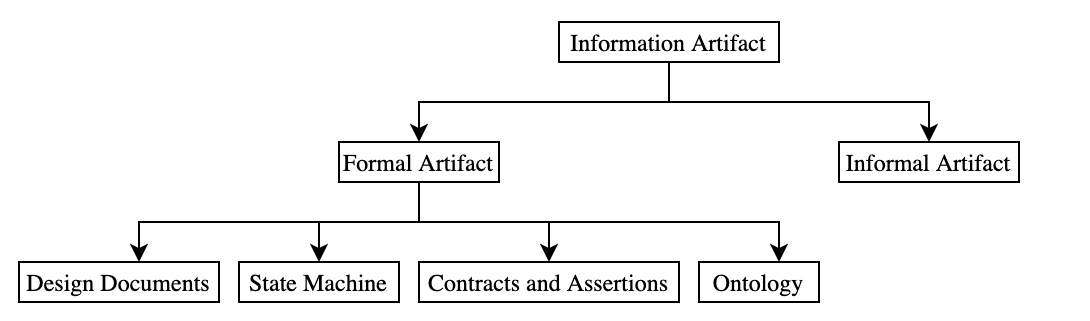}
   \caption{}
   \label{fig:informationArtifactTaxonomy}
\end{subfigure}

\begin{subfigure}{0.9\textwidth}
   \includegraphics[width=1\linewidth]{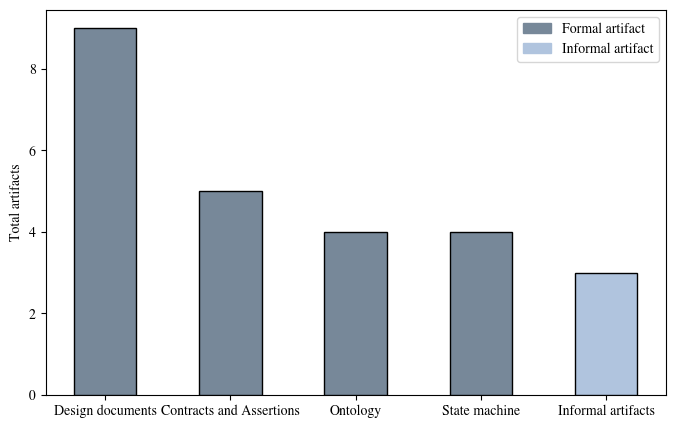}
   \caption{}
   \label{fig:informationArtifactDistribution}
\end{subfigure}

\caption{A taxonomy (a) and the distribution (b) of information artifacts that appear in the literature.}
\end{figure}




\subsubsection{Formal artifacts}
Formal artifacts express unambiguous behaviour or functionality of the system-under-test in a well-defined format. Four different forms of formal artifact have been identified in the literature: design documents, state machines, contracts and assertions, and ontologies as shown in Figure \ref{fig:informationArtifactTaxonomy}.

\emph{Design documents} - Design documents comprise a series of structured artifacts that are used in development methodologies such as Prometheus \cite{padgham2002prometheus} and Tropos \cite{bresciani2004tropos} to specify requirements, goals and scenarios for the design and development of agent-based systems. These documents contain details about the system-under-test that span the system architecture, agent communication protocol and testing goals, often taking the form of a directed graph or an AUML document. AUML is an agent-based variant of the Unified Modelling Language \cite{bauer2001agent} that is predominantly used to specify agent interaction protocols describing the semantics and structure of agent communication [P4, P9, P11]. Design documents from the Prometheus design methodology can be enriched with ``testing descriptors'' that map design variables to implementation variables, highlighting the necessary information for test input generation [P22]. Prometheus design documents have also been used as the basis for test generation in [P3, P7], and similarly, Tropos design documents have appeared in [P1, P14]. Prior to the development of the Tropos design methodology, BDI agent plans had also been used to drive test case generation [P19].

\emph{State machines} - An alternative approach is to develop state machine-based agents. In these approaches the expected behaviour of an agent-based system can be expressed using a state machine, such as an X-machine \cite{eilenberg1974automata} or stream X-machine \cite{laycock1993theory}, which enables existing automata testing theory to be applied. Regardless of their exact form, state machine specifications all contain the same basic components; states joined by transitions, where a transition corresponds to an action that moves the system from one state to another. A path through the specification corresponds to an execution of the system-under-test, and therefore a state machine specification expresses the paths through the system that may be tested \cite{chow1978testing}. A variety of different state machine information artifacts have been seen in the literature, including extended and abstract state machines [P10, P18], as well as X-machines and stream X-machines [P8, P12].

\emph{Contracts and assertions} - A different approach to formally specifying the expected behaviour of an agent-based system is the use of contracts and assertions. In contract-based testing, a contract specifies a number of pre-conditions, post-conditions, and invariants that assert the expected behaviour of a system before, after, and during execution \cite{heckel2005towards}. These conditions are a form of assertion - a logical statement that places a constraint on some state of computation \cite{korel1996assertion} - that can be used as the basis for both test case generation and an oracle. Using the formal programming language Maude \cite{clavel2002maude}, [P5] and [P6] have used rewrite rules to specify the functional behaviour of an agent-based system in terms of pre-conditions and post-conditions. The functional programming language Haskell has also been used as an executable specification to assert properties that must be satisfied during an execution and drive test case generation [P2]. Additionally, structured test scenarios containing assertion-like constructs that describe the expected outcome for a given scenario have been used as the basis for test case generation in [P17] and [P23].

\emph{Ontologies} - Ontologies are another form of formal artifact that have appeared in several of the reviewed techniques. An ontology is a collection of concepts from a given domain, including properties of the concepts and relationships between them \cite{calero2006ontologies}. In the context of agent-based models, two forms of ontology have been used as an information artifact: ``interaction ontologies'' and ``domain ontologies'' [P13]. An interaction ontology defines the concepts, relationships and interaction semantics of a model as entities and relations, providing a specification of the system that dictates how agents can interact. Whereas a domain ontology contains information specific to a domain or subject, for example, ontologies describing a production system [P20] or an environment [P21]. Using a technique known as ontology alignment, publicly available domain ontologies can be harnessed to supplement existing ontologies, enhancing the information artifact used for test case generation [P13].

\begin{table}[htbp]
\centering
\begin{tabular}{ll}
\toprule
Formal Information Artifact Type & Papers \\ \midrule
Design documents & {[}P1, P3, P4, P7, P9, P11, P14, P19, P22{]} \\
State machine & {[}P8, P10, P12, P18{]} \\
Contracts and assertions & {[}P2, P5, P6, P17, P23{]} \\
Ontologies & {[}P1, P13, P20, P21{]} \\
\bottomrule
\end{tabular}
\caption{Summary of formal information artifacts used in the evaluated literature.}
\end{table}

\subsubsection{Informal artifacts}
Informal artifacts concern qualitative features of the system-under-test that are difficult to express formally. In comparison to formal artifacts, informal artifacts appeared significantly less in the literature, as shown in Figure \ref{fig:informationArtifactDistribution}, and therefore we do not divide them further. However, it is important to distinguish between formal and informal artifacts based on the type of information they contain and the techniques that use them.

Informal artifacts typically capture qualitative features of an agent-based model such as stakeholder requirements expressed as soft goals. A soft goal is a goal that does not have a clear satisfiability definition and is often used for modelling non-functional requirements such as security, usability and flexibility \cite{padgham2008unified}. Therefore, it is not possible to definitively prove whether a soft goal has been satisfied as the threshold for satisfaction is subjective. Instead, judgements are informed by an expert opinion or a fitness function that approximates the soft goal, transforming the task of test case generation into a search problem where the objective is to find test cases that satisfy the fitness function.

Informal artifacts do not provide the same level of detail about the system-under-test as their formal counterparts, but instead, approximate desired qualities of the system-under-test that cannot be expressed formally. As a consequence, informal artifacts are generally expressed using natural language rather than a formal specification language, however, soft goals may also appear in design documents as additional information \cite{padgham2008unified}. In the reviewed literature, informal artifacts have been used to drive evolutionary approaches to test case generation, where fitness functions representing soft goals are used to continuously evolve more challenging test cases [P16, P24]. Examples of soft goals seen in the agent-based test case generation literature include reliability, efficiency and robustness [P15, P16].

\subsection{RQ2: What approaches are used to generate tests from the artifacts?}
\begin{figure}[h!]
\centering
\begin{subfigure}{\textwidth}
   \includegraphics[width=1\linewidth]{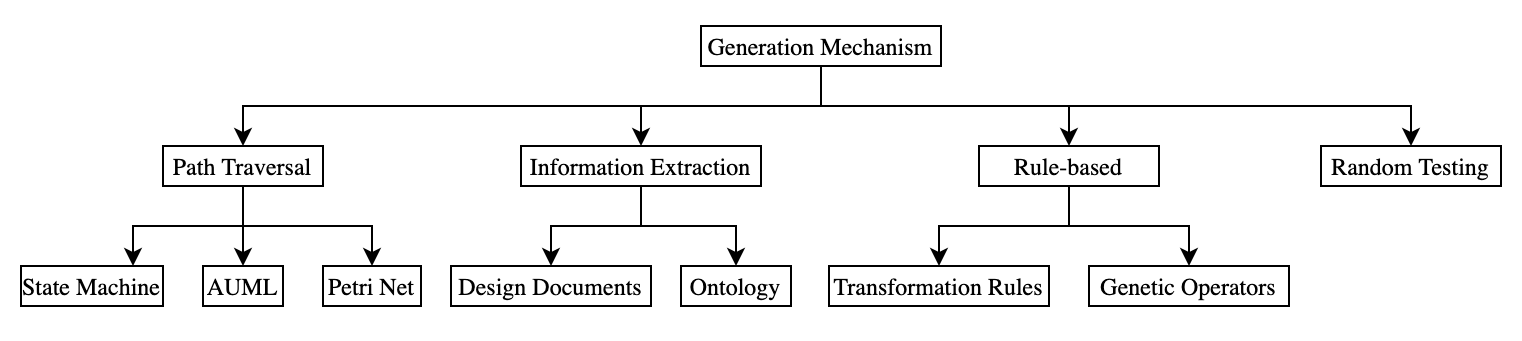}
   \caption{}
   \label{fig:generationMechanismTaxonomy}
\end{subfigure}

\begin{subfigure}{\textwidth}
   \includegraphics[width=1\linewidth]{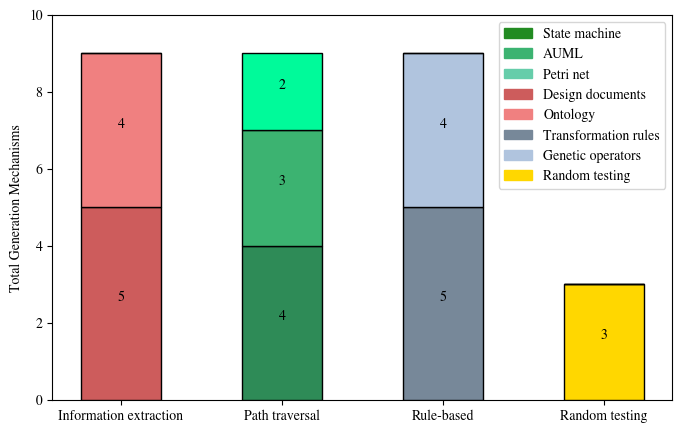}
   \caption{}
   \label{fig:generationMechanismDistribution}
\end{subfigure}

\caption{A taxonomy (a) and the distribution (b) of generation mechanisms that appear in the literature.}
\end{figure}

Four different groups of generation mechanism have been identified in the agent-based literature: path traversal, information extraction, random testing and rule-based, as shown in Figure \ref{fig:generationMechanismTaxonomy}.

%

\subsubsection{Path-traversal}
Path-traversal generation mechanisms are approaches that explore the different possible paths through a traversable representation of the system-under-test, such as a Petri net \cite{petri1962kommunikation} or a state machine \cite{gill1962introduction}, where different paths through the system correspond to different possible executions and therefore test cases. The identified path-traversal generation mechanisms can be further divided into three categories based on their subject: state machines, Petri nets and AUML.

\emph{State machine} - Several of the reviewed approaches adopt state machine-based specifications, where different paths correspond to different sequential behaviours of the system-under-test. State machines are supported by a range of automated testing techniques that make them particularly effective for specifying and testing complex systems such as agent-based models \cite{lee1996principles}. Informally, these techniques derive sequences of inputs from the specification representing paths through the system, that are used to determine whether the implementation conforms to the specification. One such technique known as the W-method \cite{chow1978testing} has been used as a generation mechanism for agent-based systems that are equipped with X-machine specifications [P8, P12]. A similar approach has also been used to generate test cases from an extended state machine specification \cite{cheng1993automatic} for the unit and pair testing of BDI agents [P10]. Test cases have also been generated using a model-checker to identify paths through a multi-agent abstract state machine that satisfy a series of data-flow coverage criteria [P18].



\emph{Petri net} - Path-traversal mechanisms also appear in the literature where Petri nets have been used as a graphical representation of the system-under-test. A Petri Net is an abstract, formal model used to describe the flow of information in systems, where dynamic components are captured by tokens that can move between states, enabling asynchronous and concurrent behaviour to be captured \cite{peterson1977petri}. Given that agents are often required to perform multiple tasks concurrently \cite{shehory2001evaluation}, Petri nets are a particularly effective tool for specifying and testing the behaviour of agent-based systems. In the reviewed literature, Petri nets have been used to represent and test the structure of BDI agent plans [P5]. Recursive coloured Petri nets \cite{haddad1999theoretical} have also been used, enabling multiple levels of abstraction of an agent-based model to be captured and accessed via recursive unfolding of nested Petri nets [P11]. To automate the generation of test cases from Petri nets, algorithms have been used that first transform the nets into a machine-readable format such as a state table, before traversing the individual paths and extracting their corresponding input sequences [P11].

\emph{AUML} - Another form of graphical representation appearing in the literature that facilitates test generation through path-traversal is AUML. Similar to Petri Nets, AUML also provides support for concurrent behaviour amongst other agent-related characteristics \cite{juneidi2010survey}, however, AUML is arguably more accessible due to its similarity to UML \cite{peres2005experiencing}. In the reviewed literature, algorithms have been presented that convert AUML sequence diagrams into a traversable format in order to identify paths through the system corresponding to test cases. For example, in an approach for testing holonic multi-agent systems \cite{gerber1999holonic}, a hierarchical AUML sequence diagram is converted into a traversable format known as a graph sequence diagram. A series of algorithms are presented to identify the different paths through the graph, before extracting the pre-conditions and post-conditions necessary to form a covering test suite [P4]. Later this approach was extended to include OCL annotations, allowing tests to identify which agent caused an error in a scenario or interaction that contains multiple agents [P9]. AUML sequence diagrams have also been translated into recursive coloured Petri nets, using its corresponding state table as the basis for test case generation [P11].

\subsubsection{Information extraction}
Information extraction generation mechanisms comprise approaches to test case generation that are driven by the extraction of test information from an information artifact, such as the extraction of variable-value pairs from design documents. In the literature, information extraction techniques have generally been applied to two different forms of information artifact: design documents and ontologies (see Figure \ref{fig:generationMechanismDistribution}).

\emph{Design documents} - As discussed in Section 4.1.1, design documents comprise a series of artifacts that are used for the design of agent-based models in agent development methodologies. For testing, useful information can be extracted from design documents and used to form test cases, such as agent goals and plans. For example, by following a structured approach, integration test suites can be derived from Tropos design documents concerning the system goals of an agent-based system [P14]. Design documents can also be enriched with test descriptors to provide additional information for test case generation, such as highlighting variables that influence a testing scenario. With this additional information, a test case specification can be extracted from design documents that highlights any initialisation procedures and relevant variables involved in a scenario that can be modified to form new test cases [P3]. Whilst this approach is only semi-automatic, a similar approach automates the extraction of variable-value pairs from Prometheus design documents. This is achieved using equivalence class partitioning and boundary value analysis to generate a minimal set of values for each influencing variable [P22]. Later, this approach was used alongside a model-based test oracle generation technique to provide automatic unit testing for agent systems [P7].


\emph{Ontology} - Another form of artifact containing useful information that may be extracted for testing is an ontology. An ontology \cite{smith2012ontology} is a formal model of knowledge representation for a given domain, detailing a set of concepts and how they are related to facilitate the exchange and organisation of knowledge \cite{moser2010ontology}. In the context of an agent-based model, ontologies are used to model a vast variety of information ranging from properties of agents and their context to interaction protocols. It follows that information extraction approaches have been proposed in the literature to utilise the information hosted by ontologies for the purpose of test case generation. One approach to ontology-based test case generation focusing on message-passing agents combines both agent interaction ontologies and domain ontologies, generating test cases in the form of messages that conform to an interaction protocol. The task of test generation then consists of generating meaningful message content that is to be sent by a \lstinline{TesterAgent} to an agent-under-test in order to provoke a particular behaviour. To generate the message content, valid and invalid inputs are generated by selecting an existing instance of a concept from the ontology that conforms to OCL constraints. Alternatively, if there is no existing concept then a new one is generated according to a series of input generation rules which will be discussed in Section 4.2.4. A technique known as ontology alignment (or ontology matching) \cite{euzenat2007ontology} is also used to expand existing ontologies by supplementing them with additional information from publicly available ontologies [P13]. Ontology-based techniques have also been used to generate test cases that focus on industry specific applications [P20] and the context of an agent-based system [P21].


\subsubsection{Random testing}
Random testing  \cite{hamlet2002random} is a well-known testing strategy where inputs are generated at random to exercise a system-under-test. As the name suggests, the generation mechanism behind random testing is the process by which random inputs are generated or selected. In comparison to other test case generation mechanisms, random testing has not received as much attention in the agent-based testing literature.

Random generation of numerical inputs is the most basic example of random testing, where in general, generation of a test case corresponds to randomly selecting a series of numerical values as input to the system-under-test. Depending on the system-under-test, the domain could be restricted to a user-defined range of numerical values or a particular set such as the natural or real numbers. Perhaps due to the challenging characteristics of an agent-based model such as communication through message-passing \cite{padgham2005developing}, richer inputs are required and therefore random generation of numerical inputs has not been used as a generation mechanism alone.

Instead, this approach has been used to support other forms of generation mechanism, such as an ontology-based technique, by selecting numerical values within a range specified as an OCL constraint to form valid and invalid inputs [P13]. Random testing can also involve sampling non-numerical, user-specified domains to generate more sophisticated test cases. For example, one approach uses random testing to sample a domain data model which defines the range and structure of messages permitted by a specific agent-interaction protocol [P1]. In addition, QuickCheck \cite{claessen2011quickcheck} - a library for random testing of program properties - has also been used to check whether a series of formally specified properties hold in an agent-based system. To achieve this, QuickCheck can automatically generate simple test inputs and provides the functionality to define custom test data generators for more complex inputs. In this work, QuickCheck is demonstrated on a simple SIR (susceptible-infectious-recovered) model \cite{bailey1975mathematical} and on a more complex model of an artificial society, SugarScape \cite{epstein1996growing}. In the former, QuickCheck generates a series of agent populations as test cases, whilst in the latter, a custom data generator is used to generate complex SugarScape environments [P2].

%

\subsubsection{Rule-based}
The final group of test generation mechanisms used in the agent-based literature concern rule-based approaches. Rule-based generation mechanisms involve a series of user-defined or automatically derived rules that describe how a test case can be manipulated to create new ones. Two types of rule-based generation mechanism have been identified in the literature: transformation rules and genetic operators.

\emph{Transformation rules} - Transformation rule-based generation mechanisms concern test generation techniques that are driven by a series of rules describing how one test case can be transformed to form a new one, by a logical transformation. User-defined transformation rules have been applied to testing agent-based models for infrastructure protection and emergency response simulation, targeting models of marine safety and security operations in particular [P17]. To this end, structured textual descriptions of example scenarios, known as vignette specifications, are modified by a library of user-defined transformation rules in order to produce different variations of the same scenarios as test cases. These transformation rules can be applied to selected variables in the vignette specification such as the number of boats involved in a scenario, their positions and their physical dimensions. Similar approaches have also used transformation rules as a generation mechanism to manipulate ``test data models'' to create new test data [P21], as well as applying them to XML definitions of an interaction protocol to form ``mock agents'' [P23].

A number of the reviewed techniques used Maude rewrite rules as a generation mechanism. In rewriting logic \cite{meseguer1992conditional}, a rewrite rule is a procedure which describes how an object can be transformed into another object. Maude \cite{clavel2002maude} is a formal language based on rewriting logic which allows users to specify components of the system-under-test and their expected behaviour as an executable specification using rewrite rules. In Maude, a rewrite rule is a local transition rule \lstinline{t => t'} that describes a valid transformation of term \lstinline{t} into another term \lstinline{t'}; for example, the rule \lstinline{rl [buy-drink] : coin => drink} describes the action of buying a drink for a coin. In the agent-based literature, Maude rewrite rules have been used as a rule-based mechanism for generating test cases alongside their expected results from a specification expressed in Maude [P5, P6].

\emph{Genetic operators} - In the context of software testing, genetic algorithms \cite{davis1991handbook} can be harnessed to evolve new test cases. Informally, a genetic algorithm is an algorithm inspired by evolution in which a solution to a specific problem, such as a test case for test case generation, is encoded as a data structure that resembles a chromosome. A series of genetic operators, namely mutation and crossover, are then applied to chromosomes repeatedly to modify them in such a way that preserves favourable features \cite{whitley1994genetic}. Mutation is the process by which the chromosomes are randomly changed, and crossover is the process in which the genetic information of two parent chromosomes are combined. To guide the algorithm towards an optimal solution, it is also necessary to define a fitness function as a measure of how well the solution attains its objective.

For testing agent-based models, genetic operators can be considered as rule-based generation mechanisms as they are applied to existing test cases to form new, potentially more challenging ones. For example, an evolutionary approach to test case generation has been proposed in the agent-based literature where the aim is to evolve test cases that are challenging with respect to soft goals such as efficiency, as discussed in Section 4.1.2. This is achieved using a multi-objective genetic algorithm, NSGA-II \cite{deb2002fast}, with multiple fitness functions that represent the soft goals. The genetic algorithm applies mutation operators and crossover with a given probability to a chromosome representing the agent-under-test's environment, producing new environments as test cases that are potentially more challenging. The approach is evaluated in an example scenario where a \lstinline{CleanerAgent} is tasked with cleaning a grid whilst maintaining a certain power level and avoiding obstacles efficiently. Test cases are generated with multiple obstructions and difficult to reach charging stations, presenting a significant challenge for the agents [P16]. This approach was later adopted to support AgentTest, a specification language for agent-based system testing [P15], before being improved with a preference-based multi-objective algorithm, r-NSGA-II, that is capable of prioritising a subset of the multiple objectives, such that generated test cases are most challenging with respect to those objectives [P24].


\begin{table}[htbp]
\centering
\begin{tabular}{ll}
\toprule
Generation Mechanism Type & Papers \\ \midrule
Path-traversal & {[}P4, P5, P8-P12, P18{]} \\
Information extraction & {[}P1, P3, P7, P13, P14, P20-P22{]} \\
Random testing & {[}P1, P2, P13{]} \\
Rule-based & {[}P1, P5, P6, P15-P17, P21, P23, P24{]} \\
\bottomrule
\end{tabular}
\caption{Summary of different generation mechanisms used in the literature.}
\end{table}

\subsection{RQ3: How are verdicts assigned to generated test cases?}
\begin{figure}[h!]
\centering
\begin{subfigure}{\textwidth}
   \centering
   \includegraphics[width=0.35\linewidth]{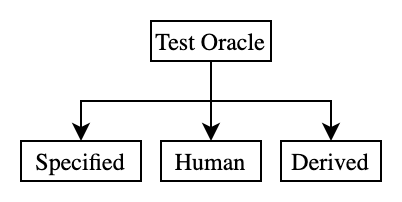}
   \caption{}
   \label{fig:testOracleTaxonomy}
\end{subfigure}

\begin{subfigure}{\textwidth}
   \centering
   \includegraphics[width=0.6\linewidth]{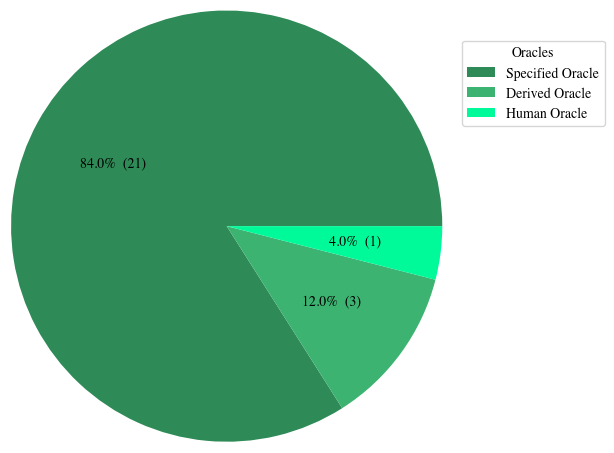}
   \caption{}
   \label{fig:testOracleDistribution}
\end{subfigure}

\caption{A taxonomy (a) and the distribution (b) of test oracles that appear in the literature.}
\end{figure}

Due to the exploratory nature of an agent-based model, the ``correct'' outcome of an execution is often unknown and it may be particularly difficult to assign a verdict to a test case. This is known formally as the test oracle problem, where the test oracle is the mechanism behind the decision. In the reviewed literature, three of the four different categories of oracle introduced by McMinn et al. \cite{mcminn2015oracle} have been identified: specified, human and derived (see Figure \ref{fig:testOracleTaxonomy}).

%

\subsubsection{Specified}
Specified test oracles are mechanisms that assign test verdicts based on a formal specification of the system-under-test \cite{mcminn2015oracle}. Formal specifications, including design documents, state machines, contracts and assertions, and ontologies, express the expected behaviour of the system-under-test under various conditions, defining what should happen and implicitly what should not. Given that the majority of the reviewed approaches involve a formal specification (discussed in Section 4.1.1), it follows that most of the approaches also include a specified oracle as shown in Figure \ref{fig:testOracleDistribution}. Whilst the oracle information is present in these approaches, several have not explained how it can be compared to the test execution in order to assign a verdict [P3, P12, P14, P19, P20, P22, P23]. On the other hand, one of the reviewed approaches focuses on the automatic generation of partial, passive test oracles from Prometheus agent design documents for BDI agent-based systems alongside a technique for test case generation [P7]. Using this approach, a selection of faults in BDI agent plans, events and beliefs can be automatically detected in an implementation of a BDI agent system, based on the expected behaviour specified in Prometheus design documents.

\subsubsection{Derived}
Derived test oracles are mechanisms that assign test verdicts based on information derived from artifacts of the system-under-test, rather than information that is explicitly defined in a formal specification \cite{mcminn2015oracle}. Whilst most of the reviewed approaches use a formal specification as an information artifact, three approaches focus on generating test cases that satisfy soft goals, or non-functional requirements, of the system-under-test that cannot be formally expressed [P15, P16, P24]. To address this issue, fitness functions have been used to approximate the fulfilment of a particular soft goal, acting as an oracle that measures whether a certain quality or non-functional requirement is satisfied by agents. Fitness functions are typically derived from qualities of the system-under-test, rather than a feature of a formal specification. For example, the efficiency of an agent cleaning a grid can be measured by a pair of fitness functions [P16]:
$$f_{power}=\frac{1}{Total \: power \: consumption}$$ $$f_{obs}=\frac{1}{Number \: of \: obstacles \: encountered}$$ In this scenario, a threshold value can be set such that any agent that exceeds the threshold is to be considered efficient, providing an approximation of correctness and a goal for test case generation.

\subsubsection{Human}
A human oracle is used where no information artifact can be used as the basis of a test oracle, and instead, a human must make the decision as to whether a test case should pass or fail using domain expertise \cite{mcminn2015oracle}. In the reviewed literature, all of the approaches use a formal or informal specification as an information artifact and thus provide at least a derived oracle. However, many of the approaches focus entirely on how test cases are generated and not how verdicts are assigned. Even in the case that the information artifact contains the information necessary to determine whether an execution is as expected, human judgement may still be required to compare the execution to the specification. Whilst this comparison is straightforward in some instances, for models with visual or otherwise complex outputs, the comparison may be subjective. In such cases, the distinction between different forms of an oracle is less clear. For example, in one state machine-based approach, a human must observe the visual output produced by the system-under-test and compare it with the expected output expressed by a stream X-machine specification [P12]. In this situation, the type of oracle used is unclear: the stream X-machine specification is a specified oracle, but it requires a subjective human decision which could itself constitute a human oracle. Whilst only one approach explicitly mentions the need for human comparison [P12], several others may also require the same treatment [P3, P14, P19, P20, P22, P23].

\begin{table}[htbp]
\centering
\begin{tabular}{ll}
\toprule
Test Oracle Type & Papers \\ \midrule
Specified & {[}P1-P14, P17-P23{]} \\
Derived & {[}P15, P16, P24{]} \\
Human & {[}P12{]} \\
\bottomrule
\end{tabular}
\caption{Summary of test oracles used in the evaluated literature.}
\end{table}

\subsection{RQ4: How is the adequacy of a generated test suite measured?}
\begin{figure}[h!]
\centering
\begin{subfigure}{0.9\textwidth}
   \centering
   \includegraphics[width=0.6\linewidth]{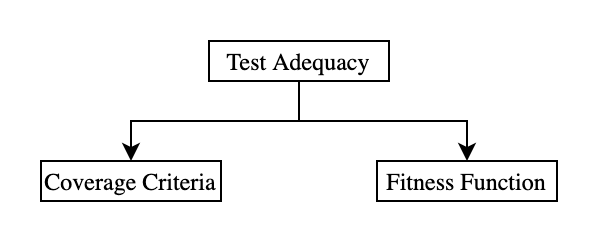}
   \caption{}
   \label{fig:testAdequacyTaxonomy}
\end{subfigure}

\begin{subfigure}{0.9\textwidth}
   \centering
   \includegraphics[width=0.5\linewidth]{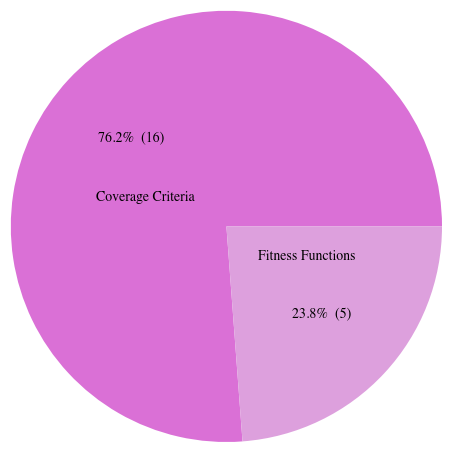}
   \caption{}
   \label{fig:testAdequacyDistribution}
\end{subfigure}

\caption{A taxonomy (a) and the distribution (b) of test adequacy criteria that appear in the literature.}
\end{figure}

An effective approach to test case generation should be guided by some measure of how \emph{good} a test suite is \cite{mathur1994empirical}, known more formally as a measure of test adequacy. In the literature, we have identified two core measures of test adequacy that have been used as the goal for test case generation: coverage criteria and fitness functions (see Figure \ref{fig:testAdequacyTaxonomy}).

%

\subsubsection{Coverage criteria}
Coverage criteria are a measure of test adequacy that focus on how thoroughly a test suite covers a specified portion of the information artifact that represents some functionality in the system-under-test. Where a state machine specification is available, test adequacy is often measured using state and transition coverage [10, 12], or alternatively, test adequacy can be measured using statement coverage, branch coverage and path coverage \cite{zhu1997software}. However, the latter criteria are not directly applicable to agent-based models as the notions of branches and paths from other programming paradigms do not translate to agent-based models clearly \cite{low1999automated}. To address this, several works have proposed coverage criteria specific to agent-based frameworks and testing approaches, including the following particularly interesting criteria: BDI artifact coverage, data-flow coverage and ontology coverage.

\emph{BDI artifact coverage} - A series of belief-desire-intention coverage criteria have been defined which draw parallels to existing coverage criteria such as branch and statement coverage, but instead apply to the equivalent BDI artifacts: plan-based and node-based criteria [P19]. The group of node-based criteria are designed to generate test cases that cover all possible states of the system, whereas the group of path-based criteria aim to test how agent plans are related to one another. Additionally, a subsumption hierarchy is defined, indicating the minimal necessary coverage criteria for generating a satisfactory test suite. Alternative definitions for coverage of BDI events, beliefs, context conditions and plans have also been given in later works [P7], including a series of state machine-based coverage criteria that are applicable when BDI agents can be modelled as an extended state machine [P10].

\emph{Data-flow coverage} - An alternative test case generation technique has been developed that focuses on data flow \cite{cavarra2011data}. In this approach, a series of data flow coverage criteria are introduced to enable data flow testing of multi-agent abstract state machines, focusing on the interaction of agent components. Data flow testing \cite{rapps1985selecting} concerns a family of testing techniques that explore how variables are defined and subsequently used in a system to reveal unreasonable data usage, such as declaring but never using a variable. Typically, data flow testing is based on the control flow graph of the system-under-test, however, an agent-based model does not have a one-to-one mapping from code to control flow graph and therefore traditional approaches to data flow testing are not directly applicable. Instead, this approach adapts traditional data flow testing techniques to handle abstract state machine representations of agent-based models, including a series of abstract state machine-specific data flow coverage criteria [P18].

\emph{Ontology coverage} - Test case generation has also been guided by the coverage of an ontology in ontology-based approaches [P13]. In this approach, concepts and instances from an ontology that have not been previously selected for test cases are prioritised to provide greater coverage of the input space, instead of focusing on a narrow subset of information. In the case that an instance must be reused in a test case, the most infrequently used instance is selected in order to maximise the diversity of the test suite. As a consequence of the thorough exploration of the input space, the fault detection capability of the generated test suite is shown to be superior to manual testing. In other approaches, coverage of concepts derived from context ontologies [P21] and coverage of the possible test cases that can be formed from a given testing scenario derived from a domain ontology [P20] have been used as measures of test data adequacy.

\subsubsection{Fitness functions}
A fitness function evaluates how well a particular solution satisfies a problem-specific objective, providing a ranking over all potential solutions \cite{whitley1994genetic}.
For the activity of test case generation, a fitness function can be used to measure the adequacy of a test case with respect to a problem-specific objective, including satisfaction of stakeholder goals such as collision avoidance [P16] or the ability to reveal known faults [P21]. In the literature, fitness functions have also been defined to capture soft goals of agent-based systems, where the adequacy of a generated test suite is measured by the fulfilment of non-functional requirements such as efficiency and reliability [P15, P16, P24]. Consequently, fitness functions have been used to guide evolutionary approaches to test case generation in which genetic operators are applied to evolve test suites that are challenging with respect to a defined objective, as described in Section 4.2.4. As part of the eCAT test case generation tool for multi-agent systems, mutation adequacy score has been used as a measure of adequacy for test suites generated using the \lstinline{EVOL-MUTATION} approach [P1]. Mutation adequacy is the fraction of injected faults that are detected by the test suite, summarising the ability of a test suite to detect known faults in the system-under-test \cite{jia2010analysis}.

\begin{table}[htbp]
\centering
\begin{tabular}{ll}
\toprule
Test Data Adequacy Criterion & Papers \\ \midrule
Coverage criteria & {[}P1, P3-P10, P12, P13, P18-P22{]} \\
Fitness function & {[}P1, P15, P16, P21, P24{]} \\
\bottomrule
\end{tabular}
\caption{Summary of test data adequacy critria used in the evaluated literature.}
\end{table}

\subsection{RQ5: What level of abstraction do the generated test cases target?}
\begin{figure}[h!]
\centering
\begin{subfigure}{0.9\textwidth}
   \centering
   \includegraphics[width=0.6\linewidth]{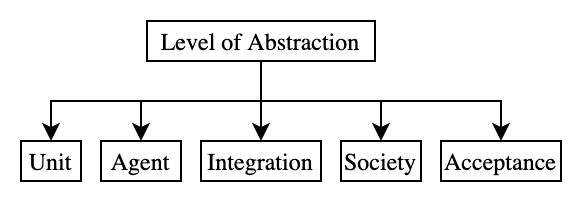}
   \caption{}
   \label{fig:levelsOfAbstractionTaxonomy}
\end{subfigure}

\begin{subfigure}{0.9\textwidth}
   \centering
   \includegraphics[width=0.9\linewidth]{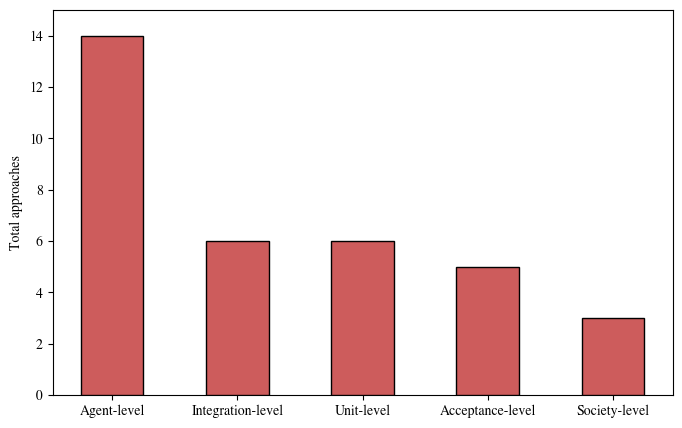}
   \caption{}
   \label{fig:levelsOfAbstractionDistribution}
\end{subfigure}

\caption{A taxonomy (a) and the distribution (b) of levels of abstraction that have been tested in the literature.}
\end{figure}

Agent-based models comprise multiple levels of abstraction, from the individual units that form a single agent up to a society composed of multiple interacting agents, each containing different functionality and testing objectives. Nguyen et al. have distinguished five different testing objectives that should be considered when testing a typical agent-based model: unit, agent, integration, society and testing \cite{nguyen2009testing} as shown in Figure \ref{fig:levelsOfAbstractionTaxonomy}.

%

\subsubsection{Unit}
The lowest level of abstraction to consider when testing an agent-based model is that of the constituent units that make up an agent, such as goals, plans and rules. Several approaches from the reviewed literature have focused on the generation of unit tests for agent-based systems alone [P7, P19, P22], whilst a further three approaches have also considered additional levels of abstraction, such as the agent-level [P2, P6, P15]. For example, the beliefs, plans and events of a BDI agent-based model have been subjected to boundary value analysis and equivalence class partitioning in order to generate variable-value pairs as test cases that exercise these units of the system-under-test [P22].

\subsubsection{Agent}
Agent-level testing concerns the integration of the units that compose an agent (covered individually in unit testing), such as goals and plans, that collectively make up the behaviour of an individual. A significant majority of the approaches from the reviewed literature focus on generating test cases that exercise agent-level behaviour alone or in addition to other levels of abstraction [P2, P5, P6, P8, P10-P13, P15, P16, P18, P21, P23, P24]. For example, using the random testing tool QuickCheck, test cases focusing on unit and agent-level behaviour have been generated for property-based testing \cite{fink1997property} of agent-based models [P2].

\subsubsection{Integration}
Integration-level testing concerns the interaction between individual agents and their environments towards cooperative behaviour. Several works have focused on both the agent-level and integration-level of interaction [P6, P10, P18, P23], where test cases target both individual agent behaviour and behaviour that is the result of agent-to-agent or agent-to-environment interactions. In addition, a pair of techniques focusing solely on integration-level testing have been proposed [P9, P14]. For example, BDI agents have been modelled as extended state machines where pair-wise interaction can be captured by the product machine composed of two individual extended state machines. As a consequence, existing testing techniques can be used to generate both agent and integration-level test cases corresponding to testing individual and cooperative behaviour [P10].

\subsubsection{Society}
Perhaps the most complicated level of abstraction to test is the society level, concerning the complex emergent behaviours that are the product of the lower levels of abstraction. Emergence in agent-based systems can be described as ``the arising of novel and coherent structures, patterns and properties during the process of self-organisation in complex systems'', as defined by Goldstein \cite{goldstein1999emergence}. Society level test case generation has received little attention with only three approaches considering this level of abstraction [P4, P6, P23]. One of the approaches targets test case generation for holonic multi-agent systems \cite{gerber1999holonic}, focusing on testing emergent behaviours that have been acquired over time in order to deal with challenging and previously unseen situations [P4]. Another approach introduces a testing framework for agent-based models that supports the testing of agent, integration and society-level testing through the generation of mock agents from XML protocol specifications that trigger a testing scenario [P23]. The third approach involves conformance testing of agent-interaction protocols specified in Maude, where the specified transition sequences are extracted and used to conduct testing at the unit, agent, integration and society levels of testing [P6].

\subsubsection{Acceptance}
The final level of abstraction to consider when testing an agent-based model is acceptance. Acceptance testing is a well-known software testing practice that involves testing whether high-level acceptance criteria are met, such as the fulfilment of user requirements \cite{miller2001acceptance}. In the reviewed literature, a couple of approaches for acceptance testing have been proposed that focus on generating alternative versions of the same user-specified scenario. One approach focuses on generating different variations of a specified marine safety and security scenario as test cases [P17], whilst the other approach is a more general testing framework that enables acceptance testing of scenarios specified as part of the Prometheus agent development methodology [P3]. In addition, evolutionary approaches to test case generation have been proposed that are suitable for requirements testing [P16, P24]. In these approaches, stakeholder goals are formulated as fitness functions which act as both the generation mechanism, as discussed in Section 4.2.4, and an oracle for evaluating whether a particular execution meets the stakeholder goal, as discussed in Section 4.3.2. This approach has also been used as a complementary testing methodology for the AgentTest specification language that covers unit, agent and acceptance level testing [P15].

\begin{table}[htbp]
\centering
\begin{tabular}{ll}
\toprule
Level of abstraction & Papers \\ \midrule
Unit & {[}P2, P6, P7, P15, P19, P22{]} \\
Agent & {[}P2, P5, P6, P8, P10-P13, P15, P16, P18, P21, P23, P24{]} \\
Integration & {[}P6, P9, P10, P14, P18, P23{]} \\
Society & {[}P4, P6, P23{]} \\
Acceptance & {[}P3, P15-P17, P24{]} \\
\bottomrule
\end{tabular}
\caption{Summary of levels of abstraction tested in the evaluated literature.}
\end{table}

\subsection{Publication trends}
In this section, the publication trends collected during the data extraction are presented, including publication venues and the results of the quality assessment. Notice that the quality assessment presented in Table \ref{tab:qualityAssessmentScores} does not include any literature with $quality \leq 1.5$, as any study below this threshold would be excluded due to a perceived lack of quality in accordance to the quality assessment procedure discussed in Section 3.6.

\begin{table}[!htbp]
\footnotesize
\centering
\begin{tabular}{lc}
\toprule
Venue & Count \\ \midrule
International Conference on Autonomous Agents and Multi-Agent Systems & 4 \\
International Workshop on Agent-Oriented Software Engineering & 2 \\
IEEE/ACS International Conference on Computer Systems and Applications & 2 \\
International Conference on Software Engineering and Knowledge Engineering & 1 \\
International Conference on Informatics in Control, Automation and Robotics & 1 \\
International Conference on Agents and Artificial Intelligence & 1 \\
IEEE Transactions on Software Engineering & 1 \\
IEEE International Conference on Software Maintenance & 1 \\
Summer Simulation Conference & 1 \\
Engineering Applications of Artificial Intelligence & 1 \\
Science and Information Conference & 1 \\
Federated Conference on Computer Science and Information Systems  & 1 \\
International Conference on Networking and Advanced Systems & 1 \\
International Symposium on Search Based Software Engineering & 1 \\
Asia-Pacific Software Engineering Conference  & 1 \\
Journal of Computer Science & 1 \\
Neurocomputing & 1 \\
Security Informatics & 1 \\
Formal Aspects of Computing & 1 \\
\bottomrule
\end{tabular}
\caption{Publication venues of reviewed work.}
\label{tab:venues}
\end{table}

As shown in Table \ref{tab:venues}, the reviewed literature has been published in a variety of conferences and journals spanning multiple subject areas. Whilst this demonstrates that agent-based models and test case generation are multidisciplinary research topics, it also highlights the lack of an established publication channel.


\begin{table}[!htbp]
\footnotesize
\centering
\begin{tabular}{llllll}
\toprule
ID & QA1 & QA2 & QA3 & QA4 & Total \\ \midrule
{[}P1{]} & 1 & 1 & 0 & 1 & 3 \\
{[}P2{]} & 1 & 1 & 0.5 & 1 & 3.5 \\
{[}P3{]} & 1 & 1 & 0.5 & 0.5 & 3 \\
{[}P4{]} & 1 & 1 & 1 & 1 & 4 \\
{[}P5{]} & 1 & 1 & 0.5 & 0.5 & 3 \\
{[}P6{]} & 1 & 1 & 0.5 & 0.5 & 3 \\
{[}P7{]} & 1 & 1 & 0.5 & 1 & 3.5 \\
{[}P8{]} & 1 & 1 & 0.5 & 0.5 & 3 \\
{[}P9{]} & 1 & 0.5 & 1 & 0 & 2.5 \\
{[}P10{]} & 1 & 1 & 1 & 0 & 3 \\
{[}P11{]} & 1 & 1 & 0.5 & 0.5 & 3 \\
{[}P12{]} & 1 & 1 & 0.5 & 0.5 & 3 \\
{[}P13{]} & 1 & 1 & 0.5 & 0.5 & 3 \\
{[}P14{]} & 1 & 0.5 & 0.5 & 0.5 & 2.5 \\
{[}P15{]} & 1 & 1 & 0.5 & 0.5 & 3 \\
{[}P16{]} & 1 & 1 & 1 & 0.5 & 3.5 \\
{[}P17{]} & 1 & 1 & 0.5 & 1 & 3.5 \\
{[}P18{]} & 1 & 1 & 0.5 & 0.5 & 3 \\
{[}P19{]} & 1 & 1 & 0.5 & 0.5 & 3 \\
{[}P20{]} & 0.5 & 0.5 & 0.5 & 0.5 & 2 \\
{[}P21{]} & 0.5 & 0.5 & 0.5 & 0.5 & 2 \\
{[}P22{]} & 1 & 1 & 0.5 & 0.5 & 3 \\
{[}P23{]} & 1 & 1 & 1 & 0.5 & 3.5 \\
{[}P24{]} & 1 & 1 & 0.5 & 0.5 & 3 \\
\bottomrule
\end{tabular}
\caption{Quality assessment scores of the reviewed literature.}
\label{tab:qualityAssessmentScores}
\end{table}


As shown in Table \ref{tab:qualityAssessmentScores}, the majority of the reviewed publications scored highly in QA1 and QA2 as a result of having well-defined aims that were clearly met. However, many of the publications did not include a thorough evaluation of the proposed technique and did not include sufficient details for replication. Consequently, studies did not score as highly in QA3 and QA4 in general.

\clearpage
\section{Discussion}
This section provides a collective discussion of the reviewed literature, highlighting the advantages and disadvantages of the reviewed techniques, interesting observations from each research questions, and future research directions.

\subsection{Advantages and disadvantages}
In response to RQ1 and RQ2, we reviewed a range of information artifacts and generation mechanisms, each having different advantages and disadvantages that make it more or less suitable for particular applications. In Tables \ref{tab:advantagesAndDisadvantagesIA} and \ref{tab:advantagesAndDisadvantagesGM}, we summarise the relevant advantages and disadvantages of the reviewed information artifacts and generation mechanisms respectively.

{
\renewcommand{\arraystretch}{1.2}
\begin{table}[htb]
\footnotesize
\begin{tabularx}{\linewidth}{XXX}
\toprule
Information Artifact & Advantage & Disadvantage  \\ \midrule

Design documents & 
Widely used in agent-oriented development and therefore applicable to many existing systems \cite{padgham2013model}. &
Requires the user to follow particular design methodologies which are time consuming to learn \cite{hadar2010empirical}. \\

State machines &
Supports the use of a range of established test generation techniques \cite{kalaji2009generating}. &
Requires the user to learn how to formally specify state machines \cite{friedman2002projected}. \\

Contracts and assertions &
Users can specify behaviour independent of the implementation i.e. what the model should do, not how the model should do it \cite{thaler2019show}. &
No guarantee that a test case will be generated that, upon execution, covers
a specified contract or assertion \cite{leitner2007contract}. \\

Ontologies &
Supported by many popular agent development platforms including JADE \cite{nguyen2008experimental}. &
Converting domain expertise into an ontology is difficult and time consuming \cite{tan2019lessons}. \\
\bottomrule

\end{tabularx}
\caption{Advantages and disadvantages of the different information artifacts}
\label{tab:advantagesAndDisadvantagesIA}
\end{table}
\clearpage

\begin{table}[h!]
\footnotesize
\begin{tabularx}{\linewidth}{XXX}
\toprule
Generation Mechanism & Advantage & Disadvantage  \\ \midrule

Path-traversal (State machine) &
Under specific design-for-test conditions, certain techniques can guarantee correctness of the implementation \cite{eleftherakis2011methodology}. &
Suffers from state-explosion problem, particularly in concurrent systems \cite{friedman2002projected}. \\

Path-traversal (Petri net) &
Provides a natural representation for concurrent behaviour \cite{kissoum2008recursive}, a defining feature of agent-based models. &
Suffers from the complexity problem. That is, they quickly become too large to analyse for even modest-size systems \cite{murata1989petri}.\\

Path-traversal (AUML) &
Effective for generating test cases that exercise agent interactions and scenarios \cite{kissoum2008recursive, dehimi2019novel}. &
Agents are represented as objects in AUML, limiting the ability to express and test proactive behaviour \cite{gonccalves2015mas}. \\

Information extraction (Design documents) &
Supports class partitioning and boundary value analysis to efficiently achieve a particular level of coverage \cite{zhang2009automated}. &
Additional testing information must be added manually to design documents to enable test case generation \cite{zhang2009automated, thangarajah2011scenarios}. \\

Information extraction (Ontology) &
Supports ontology alignment to expand and diversify the available testing information \cite{nguyen2008experimental}. &
Ontology-based approaches require greater effort than manual testing to set up initially, but are more efficient in the long term \cite{moser2010ontology}. \\

Random testing &
Capable of generating large amounts of test cases for a relatively small cost \cite{arcuri2011random}. &
Requires manual intervention or custom data generators to handle complex, non-numerical inputs \cite{thaler2019show}. \\

Rule-based (Transformation rules) &
Enables the automatic generation of domain-relevant test cases using reusable transformation rules \cite{shahir2012generating}. &
Transformation rules must be manually specified based on domain expertise \cite{shahir2012generating}, often using a formal language such as Maude \cite{mokhati2013novel}. \\

Rule-based (Genetic operators) &
Enables the generation of test cases that are challenging with respect to a specific user-defined testing objective \cite{nguyen2012evolutionary}. &
Performance is dependent on how accurately the fitness function captures the testing objective \cite{nguyen2012evolutionary}. \\
\bottomrule

\end{tabularx}
\caption{Advantages and disadvantages of the different generation mechanisms}
\label{tab:advantagesAndDisadvantagesGM}
\end{table}
}

\subsection{Discussion of results}
In response to RQ1, two different groups of information artifacts were identified: formal and informal. The results to RQ1 showed that significantly more approaches use formal information artifacts than informal information artifacts as the basis for test generation. As a consequence, the majority of the reviewed approaches focus on well-defined, functional properties of an agent-based system that can be verified using existing, well-established testing techniques. For example, where state machines have been used as a formal specification of the system-under-test, the W-method may be used to generate test cases. However, a state machine does not support the specification and testing of non-functional, qualitative properties intrinsic to agent-based models such as efficiency and reliability. Conversely, only a small number of approaches have used informal information artifacts to drive test case generation, that instead focus on approximating soft goals and testing whether they are fulfilled. \\

\noindent\fbox{%
    \parbox{\linewidth}{%
        \textbf{Observation 1:} The majority of reviewed information artifacts do not facilitate specification and testing of non-functional requirements such as soft goals.
    }%
} \\[0.2cm]

In response to RQ2, four groups of generation mechanism were found: information extraction, path traversal, rule-based and random testing. Following the outcome of RQ1, most of these generation mechanisms harness information expressed by a formal information artifact, such as a Petri net, state machine or ontology, to form new test cases that test different behaviour. Whilst path-traversal, information extraction and rule-based mechanisms were extensively used, only a small number of approaches used random testing as a mechanism for test case generation. \\

\noindent\fbox{%
    \parbox{\linewidth}{%
        \textbf{Observation 2:} Information extraction, path traversal and rule-based mechanisms have been used as techniques for generating test cases that exercise functional requirements.
    }%
} \\[0.2cm]

We found that in the few approaches where informal artifacts had been used, genetic algorithms had been adopted as a solution to generate test cases that satisfy a non-functional requirement. In these approaches, the genetic operators known as mutation and crossover are repeatedly applied to seed test cases to form new, potentially more challenging tests with respect to a non-functional requirement expressed as a fitness function. \\

\noindent\fbox{%
    \parbox{\linewidth}{%
        \textbf{Observation 3:} Genetic algorithms have been used as a technique for evolving test cases that exercise non-functional requirements.
    }%
} \\[0.2cm]

In response to RQ3, three forms of test oracle were discussed: specified, derived, and human. Due to most approaches involving some form of specification as an information artifact, the vast majority of approaches involved a specified oracle. Where informal specifications of non-functional requirements were used, oracles were generally derived from the system-under-test or domain knowledge in the form of fitness functions that approximate soft goals. One approach also required a human oracle to judge whether the visual output of an agent-based model matched the specification. This research question proved particularly difficult to answer as most of the reviewed literature did not discuss the test oracle in great detail, neglecting information about how the outcome of a test execution is compared to the specification in order to assign a verdict. Whilst in some cases this is a trivial comparison, in others, it is not so straightforward and it appears that domain expertise are needed to compare the actual execution to the expected execution, raising questions over whether the oracle should be considered specified or human. Therefore, it is necessary to communicate all details of the oracle. \\

\noindent\fbox{%
    \parbox{\linewidth}{%
        \textbf{Observation 4:} Most of the reviewed approaches do not provide sufficient details regarding the test oracle.
    }%
} \\[0.2cm]

In response to RQ4, two forms of test data adequacy criterion were identified: coverage criteria and fitness functions. Where a formal information artifact has been used to specify the behaviour of the system-under-test, we found that a variety of different agent-based coverage metrics were used as a measure of test data adequacy. For example, in some BDI agent systems, test suites have been created with the aim of covering various different artifacts specific to the BDI agent architecture such as agent plans. \\

\noindent\fbox{%
    \parbox{\linewidth}{%
        \textbf{Observation 5:} The adequacy of a functional test suite is generally measured by some form of agent-based coverage.
    }%
} \\[0.2cm]

Where an informal information artifact has been used, fitness functions capturing application-specific soft goals have generally been used as a measure of test adequacy, such that an adequate test suite is one that poses a significant challenge with respect to the non-functional requirement in question. Therefore, the goal of techniques driven by informal artifacts is not to cover a particular component of the system-under-test, but to develop test cases that force the system-under-test into extreme conditions under which soft goals such as efficiency become increasingly difficult to satisfy. \\

\noindent\fbox{%
    \parbox{\linewidth}{%
        \textbf{Observation 6:} The adequacy of a non-functional test case is generally measured using a fitness function that represents fulfilment of the non-functional requirement.
    }%
} \\[0.2cm]

In response to RQ5, we classified each of the reviewed approaches according to the level of abstraction that the generated test suites target: unit, agent, integration, society, and acceptance. We found that the significant majority of approaches target the agent level of testing, generating test cases that target the functionality of individual agents. Several approaches also targeted the unit, integration and acceptance levels of testing. However, the society level of testing has received far less attention than the other levels of abstraction. \\

\noindent\fbox{%
    \parbox{\linewidth}{%
        \textbf{Observation 7:} Most of the reviewed approaches target the agent-level of abstraction, whilst the society-level has received the least attention.
    }%
} \\[0.2cm]

After conducting the quality assessment, we found that the quality of the reviewed approaches was generally high, with most of the reviewed literature scoring highly in QA1 and QA2 reflecting clear definition and fulfilment of aims. However, we found that in most cases there were insufficient details to replicate the approach, and in particular, only a few approaches were evaluated on ``real-world'' agent-based systems.

Whilst the majority of techniques were demonstrated on toy examples, one technique was evaluated thoroughly on a complex production automation system \cite{moser2010ontology}. The evaluation focused on the feasibility, cost-reducing capability, and coverage of possible scenarios achieved using an ontology-based test case generation technique. The authors demonstrated that in the long-term, ontology-based test case generation outperforms manual test case generation in all dimensions.  Given the success of this technique in an industrial context, future work should also consider real-world case studies for evaluation in order to demonstrate their potential in industry. \\



\noindent\fbox{%
    \parbox{\linewidth}{%
        \textbf{Observation 8:} Evaluation is rarely performed on real-world systems that include typical agent-based properties such as concurrency and non-determinism.
    }%
} \\[0.2cm]

Most of the reviewed literature did not include an experimental evaluation. Instead, most techniques were demonstrated on example systems ranging from bibliographic management systems to implementations of agent interaction protocols. However, these did not incorporate a controlled comparison with other techniques. In the few papers that did conduct an experimental evaluation, mutation testing was generally applied to an example system. These works report the mutation score or fault detection rate of their test generation technique with respect to a number of known or seeded faults.

Whilst mutation testing provides a measure of fault detection capability, multiple confounding factors must be considered in order to provide a robust comparison between techniques \cite{papadakis2019mutation}. In particular, the results of mutation testing depend on the introduced faults and the system-under-test. Therefore, to achieve a fair comparison, techniques should be compared in terms of their ability to reveal faults in identical systems. To this end, future work would benefit from a database of benchmark agent-based systems seeded with faults, similar to the Defects4J framework in Java \cite{just2014defects4j}. \\

\noindent\fbox{%
    \parbox{\linewidth}{%
        \textbf{Observation 9:} Existing techniques are difficult to compare due to a lack of consistent experimental evaluation methodology.
    }%
} \\[0.2cm]

\subsection{Impact on research and practice}
The findings of this literature review have several implications for both research and practice concerning agent-based models:

\begin{enumerate}
  \item Testing of non-functional requirements needs greater attention from researchers. There are comparatively few test case generation techniques that support the specification and testing of non-functional requirements in agent-based models. This is a significant limitation as agent-based models often have soft goals as requirements, such as reliability and safety, that the majority of approaches cannot test.
  \item To advance the reproducibility and comparability of techniques, researchers should provide greater detail about the test oracle. Without full details of the test oracle, the reader is left to make implicit assumptions about the testing procedure, making the comparison of research difficult \cite{staats2011programs}.
  \item Additionally, future work would be more comparable if experimental evaluations used the same subject systems. To that end, we recommend that a collection of faulty agent-based models, similar to the Defects4J framework in Java \cite{just2014defects4j}, should be developed to be used as a baseline for comparing agent-based testing techniques.
  \item Due to a lack of rigorous experimental evaluation, it is currently unclear how effective the reviewed approaches would be in practice. In order to achieve more rigorous and realistic evaluations, future work should consider using real-world case studies as examples, such as the complex automation system studied by Moser et al. \cite{moser2010ontology}.
 \item In RQ1, we identified a number of forms of information artifact. A potential barrier to their adoption in industry is the time required to learn and construct them. Future work should investigate methodologies for developing user-friendly artifacts in a timely manner so as to encourage wider adoption.
 \item Only one of the reviewed test case generation techniques covered both functional and non-functional test input generation. Therefore, practitioners will usually have to adopt multiple techniques in order to generate test cases that exercise both functional and non-functional requirements. To address this, future tools would benefit from providing both functional and non-functional test case generation.
 \item Most of the reviewed test case generation techniques are only applicable to a particular form of agent-based model or development methodology. Future work should consider techniques that could be applied more generally to any form of agent-based model, without requiring significant additional effort. We believe that this is one of the necessary steps towards adoption in industry.
 
\end{enumerate}

\subsection{Threats to validity}
Following Kitchenham and Charter's guidelines for conducting systematic literature reviews, we have considered potential threats to validity that may affect the results presented in this systematic literature review. We discuss the following categories of threats to validity defined by Zhu et al. \cite{zhou2016map}: construct validity, internal validity, external validity and conclusion validity.

\subsubsection{Construct validity}
Construct validity primarily concerns the review protocol and how inclusive the selected search string, data sources and screening criteria were with respect to the research questions. To minimise threats to construct validity, we included a validation procedure for each of the components of the review protocol. For example, when designing the search string the authors each selected a sample of three papers that they believed the search should retrieve. The search string was then iteratively improved until it could retrieve all of the selected literature. However, it is still possible that some relevant literature may have been missed by the search due to the choice of search terms, data sources and screening criteria.

\subsubsection{Internal validity}
Internal validity concerns the trustworthiness of the cause-effect relationship between the results and the review methodology. In this review, the search procedure was primarily conducted by a single researcher, and therefore the results may be subject to selection bias. To reduce this, a validation procedure was conducted where a randomly selected collection of papers were reviewed by the other authors, such that each paper was reviewed by both the first author and at least one other author. Out of a set of 32 papers, 84\% yielded the same verdict by both authors, and after group discussion over the studies yielding a disagreeing verdict, 90\% of the other authors' verdicts were in agreement with the primary authors. Furthermore, the search procedure has been conducted to favour false positives over false negatives, such that a study was only rejected if the author was certain that it was irrelevant according to the screening criteria. In the case of any doubt, a paper would be moved forwards to the next stage, and if after a full-text review the verdict of a paper was still unclear, it would be reviewed by all authors independently and discussed until a group consensus was reached.

\subsubsection{External validity}
External validity concerns the applicability of the results to a given domain, which in our case is testing for agent-based models. Due to the lack of an agreed-upon definition of an agent-based model, it is possible that some of the reviewed approaches do not conform to a particular definition of an agent-based model. To minimise this threat, we provided an inclusive definition of an agent-based model in Section 2.1 and designed our review protocol accordingly. Furthermore, we carefully designed a taxonomy to group similar responses to each research question under a generalised category, providing results that generalise to various forms of agent-based model regardless of their exact details.

\subsubsection{Conclusion validity}
Conclusion validity concerns the reproducibility of the results. In order to maximise the reproducibility of our results, we followed Kitchenham and Charter's guidelines for systematic literature reviews to provide details including the choice of search string, data sources and screening criteria. We also introduced a taxonomy based on the reviewed literature to provide a more structured approach for answering the qualitative research questions. However, there remains subjectivity over the classification of an approach according to the taxonomy and therefore we cannot guarantee that upon replication the exact same conclusions would be met.

\section{Conclusion}
With the need for drastic improvements in model validation and testing being highlighted recently \cite{wynants2020prediction, squazzoni2020computational}, it is clear that more effective and accessible techniques must be developed for testing agent-based models. In this review, we have collected, organised, and presented the literature on test case generation for agent-based models, providing insights into the current state-of-the-art for test case generation in agent-based models. The systematic literature review was conducted using five digital libraries, yielding a total of 464 papers from which 24 relevant studies were identified. We used the relevant studies to answer five research questions related to the key aspects of test case generation: the information artifact, the generation mechanism, the test oracle, the test data adequacy criterion, and the level of abstraction tested.

The results showed that 88\% of the reviewed approaches focus on generating test cases that test functional requirements, and conversely, that only 12\% of approaches focus on testing non-functional requirements. This statistic suggests that testing of soft goals is not well supported by the majority of the existing approaches, limiting the potential of these techniques in practice. We also found that the majority of the techniques could be classified as information extraction, path-traversal or rule-based in terms of the mechanism used to create test cases. Furthermore, we identified a number of agent-based coverage criteria used to measure the adequacy of functional test suites and described the use of fitness functions to measure the adequacy of non-functional test suites.

This review has also revealed a lack of information regarding test oracles in the agent-based literature. To increase reproducibility and comparability of results and techniques, the procedure for assigning a verdict to a test case based on the provided information must be clearly communicated. It is not enough to state that ``the expected outcome can then be compared to the actual outcome''. Most of the reviewed approaches were also found to generate test cases that exercise agent-level behaviour, whereas the emergent societal behaviours inherent to agent-based systems have been largely ignored. Similarly, we found that evaluation of the reviewed approaches were often conducted using simplistic examples that do not exhibit properties typical of an agent-based system that would make test case generation considerably more difficult. Therefore, future techniques should consider society-level testing and use more realistic examples for evaluation that exhibit challenging agent-based properties such as non-determinism and concurrency.

\section*{Declaration of competing interest}
The authors declare that they have no known competing financial interests or personal relationships that could have appeared to influence the work reported in this paper.

\section*{Acknowledgements}
Walkinshaw and Hierons are funded by the EPSRC CITCOM grant EP/T030526/1.

\appendix
\section*{Appendix. Research papers evalutated in the review}
\begin{xltabular}{\linewidth}{lX}
{[}P1{]} & Nguyen, C.D., Perini, A. and Tonella, P., 2008, May. eCAT: a tool for automating test cases generation and execution in testing multi-agent systems. In Proceedings of the 7th international joint conference on Autonomous agents and multiagent systems: demo papers (pp. 1669-1670). \\
{[}P2{]} & Thaler, J. and Siebers, P.O., 2019, July. Show me your properties: the potential of property-based testing in agent-based simulation. In Proceedings of the 2019 Summer Simulation Conference (p. 1). Society for Computer Simulation International. \\
{[}P3{]} & Thangarajah, J., Jayatilleke, G. and Padgham, L., 2011. Scenarios for system requirements traceability and testing. In Autonomous Agents and MultiAgent Systems (pp. 285-292). IFAAMAS. \\
{[}P4{]} & Dehimi, N.E.H., Mokhati, F. and Badri, M., 2015. Testing HMAS-based applications: An ASPECS-based approach. Engineering Applications of Artificial Intelligence, 46, pp.232-257. \\
{[}P5{]} & Kissoum, Y. and Sahnoun, Z., 2007, May. A formal approach for functional and structural test case generation in multi-agent systems. In 2007 IEEE/ACS International Conference on Computer Systems and Applications (pp. 76-83). IEEE. \\
{[}P6{]} & Mokhati, F., Badri, M. and Zerrougui, S., 2013, October. A novel conformance testing technique for Agent Interaction Protocols. In 2013 Science and Information Conference (pp. 485-495). IEEE. \\
{[}P7{]} & Padgham, L., Zhang, Z., Thangarajah, J. and Miller, T., 2013. Model-based test oracle generation for automated unit testing of agent systems. IEEE Transactions on Software Engineering, 39(9), pp.1230-1244. \\
{[}P8{]} & Eleftherakis, G., Kefalas, P. and Kehris, E., 2011, September. A methodology for developing component-based agent systems focusing on component quality. In 2011 Federated Conference on Computer Science and Information Systems (FedCSIS) (pp. 561-568). IEEE. \\
{[}P9{]} & Dehimi, N.E.H. and Mokhati, F., 2019, June. A Novel Test Case Generation Approach based on AUML sequence diagram. In 2019 International Conference on Networking and Advanced Systems (ICNAS) (pp. 1-4). IEEE. \\
{[}P10{]} & Zheng, M. and Alagar, V.S., 2005, December. Conformance testing of BDI properties in agent-based software. In 12th Asia-Pacific Software Engineering Conference (APSEC'05) (pp. 8-pp). IEEE. \\
{[}P11{]} & Kissoum, Y. and Sahnoun, Z., 2008, March. A Recursive Colored Petri Nets semantics for AUML as base of test case generation. In 2008 IEEE/ACS International Conference on Computer Systems and Applications (pp. 785-792). IEEE. \\
{[}P12{]} & Sakellariou, I., Dranidis, D., Ntika, M. and Kefalas, P., 2015, January. Stream X-Machines for Agent Simulation Test Case Generation. In International Conference on Agents and Artificial Intelligence (pp. 37-57). Springer, Cham. \\
{[}P13{]} & Nguyen, C.D., Perini, A. and Tonella, P., 2008, May. Experimental evaluation of ontology-based test generation for multi-agent systems. In International Workshop on Agent-Oriented Software Engineering (pp. 187-198). Springer, Berlin, Heidelberg. \\
{[}P14{]} & Houhamdi, Z. and Athamena, B., 2011. Structured integration test suite generation process for multi-agent system. Journal of Computer Science, 7(5), p.690. \\
{[}P15{]} & Babac, M.B. and Jevtić, D., 2014. AgentTest: A specification language for agent-based system testing. Neurocomputing, 146, pp.230-248. \\
{[}P16{]} & Nguyen, C.D., Miles, S., Perini, A., Tonella, P., Harman, M. and Luck, M., 2012. Evolutionary testing of autonomous software agents. Autonomous Agents and Multi-Agent Systems, 25(2), pp.260-283. \\
{[}P17{]} & Shahir, H.Y., Glässer, U., Farahbod, R., Jackson, P. and Wehn, H., 2012. Generating test cases for marine safety and security scenarios: a composition framework. Security Informatics, 1(1), p.4. \\
{[}P18{]} & Cavarra, A., 2011. A data-flow approach to test multi-agent ASMs. Formal aspects of computing, 23(1), pp.21-41. \\
{[}P19{]} & Low, C.K., Chen, T.Y. and Rónnquist, R., 1999. Automated test case generation for BDI agents. Autonomous Agents and Multi-Agent Systems, 2(4), pp.311-332. \\
{[}P20{]} & Moser, T., Dürr, G. and Biffl, S., 2010. Ontology-Based Test Case Generation For Simulating Complex Production Automation Systems. In SEKE (pp. 478-482). \\
{[}P21{]} & Szatmári, Z., Oláh, J. and Majzik, I., 2011, July. Ontology-based Test Data Generation using Metaheuristics. In ICINCO (2) (pp. 217-222). \\
{[}P22{]} & Zhang, Z., Thangarajah, J. and Padgham, L., 2009, May. Automated testing for intelligent agent systems. In International Workshop on Agent-Oriented Software Engineering (pp. 66-79). Springer, Berlin, Heidelberg. \\
{[}P23{]} & Coelho, R., Cirilo, E., Kulesza, U., von Staa, A., Rashid, A. and Lucena, C., 2007, October. Jat: A test automation framework for multi-agent systems. In 2007 IEEE International Conference on Software Maintenance (pp. 425-434). IEEE. \\
{[}P24{]} & Kalboussi, S., Bechikh, S., Kessentini, M. and Said, L.B., 2013, August. Preference-based many-objective evolutionary testing generates harder test cases for autonomous agents. In International Symposium on Search Based Software Engineering (pp. 245-250). Springer, Berlin, Heidelberg. \\
\end{xltabular}


\bibliographystyle{unsrt}
\bibliography{bibliography.bib}

\begin{thebibliography}{100}

\bibitem{liu2020interbank}
Anqi Liu, Mark Paddrik, Steve~Y Yang, and Xingjia Zhang.
\newblock Interbank contagion: An agent-based model approach to endogenously
  formed networks.
\newblock {\em Journal of Banking \& Finance}, 112:105191, 2020.

\bibitem{tracy2018agent}
Melissa Tracy, Magdalena Cerd{\'a}, and Katherine~M Keyes.
\newblock Agent-based modeling in public health: current applications and
  future directions.
\newblock {\em Annual review of public health}, 39:77--94, 2018.

\bibitem{flaxman2020report}
Seth Flaxman, Swapnil Mishra, Axel Gandy, H~Unwin, Helen Coupland, T~Mellan,
  Harisson Zhu, T~Berah, J~Eaton, P~Perez~Guzman, et~al.
\newblock Report 13: Estimating the number of infections and the impact of
  non-pharmaceutical interventions on covid-19 in 11 european countries.
\newblock 2020.

\bibitem{panovska2020determining}
Jasmina Panovska-Griffiths, Cliff Kerr, Robyn~Margaret Stuart, Dina Mistry,
  Daniel Klein, Russell~M Viner, and Chris Bonell.
\newblock Determining the optimal strategy for reopening schools, work and
  society in the uk: balancing earlier opening and the impact of test and trace
  strategies with the risk of occurrence of a secondary covid-19 pandemic wave.
\newblock {\em medRxiv}, 2020.

\bibitem{kerr2020covasim}
Cliff~C Kerr, Robyn~M Stuart, Dina Mistry, Romesh~G Abeysuriya, Gregory Hart,
  Katherine Rosenfeld, Prashanth Selvaraj, Rafael~C Nunez, Brittany Hagedorn,
  Lauren George, et~al.
\newblock Covasim: an agent-based model of covid-19 dynamics and interventions.
\newblock {\em medRxiv}, 2020.

\bibitem{ramler2006economic}
Rudolf Ramler and Klaus Wolfmaier.
\newblock Economic perspectives in test automation: balancing automated and
  manual testing with opportunity cost.
\newblock In {\em Proceedings of the 2006 international workshop on Automation
  of software test}, pages 85--91, 2006.

\bibitem{mcminn2004search}
Phil McMinn.
\newblock Search-based software test data generation: a survey.
\newblock {\em Software testing, Verification and reliability}, 14(2):105--156,
  2004.

\bibitem{godefroid2008automated}
Patrice Godefroid, Michael~Y Levin, David~A Molnar, et~al.
\newblock Automated whitebox fuzz testing.
\newblock In {\em NDSS}, volume~8, pages 151--166, 2008.

\bibitem{prasanna2005survey}
M~Prasanna, S~Sivanandam, R~Venkatesan, and R~Sundarrajan.
\newblock A survey on automatic test case generation.
\newblock {\em Academic Open Internet Journal}, 15(6), 2005.

\bibitem{anand2013orchestrated}
Saswat Anand, Edmund~K Burke, Tsong~Yueh Chen, John Clark, Myra~B Cohen,
  Wolfgang Grieskamp, Mark Harman, Mary~Jean Harrold, Phil Mcminn, Antonia
  Bertolino, et~al.
\newblock An orchestrated survey of methodologies for automated software test
  case generation.
\newblock {\em Journal of Systems and Software}, 86(8):1978--2001, 2013.

\bibitem{kanewala2014testing}
Upulee Kanewala and James~M Bieman.
\newblock Testing scientific software: A systematic literature review.
\newblock {\em Information and software technology}, 56(10):1219--1232, 2014.

\bibitem{russell2002artificial}
Stuart Russell and Peter Norvig.
\newblock Artificial intelligence: a modern approach.
\newblock 2002.

\bibitem{luck2005agent}
Michael Luck, Peter McBurney, Onn Shehory, and Steve Willmott.
\newblock {\em Agent technology: computing as interaction (a roadmap for agent
  based computing)}.
\newblock University of Southampton, 2005.

\bibitem{obr2020coronavirus:online}
Office for Budget~Responsibility.
\newblock Coronavirus analysis.
\newblock \url{http://obr.uk/coronavirus-analysis/}, April 2020.
\newblock (Accessed on 05/19/2020).

\bibitem{mrcideco79:online}
Covid-19 covidsim model.
\newblock \url{https://github.com/mrc-ide/covid-sim}, 2020.
\newblock (Accessed on 05/19/2020).

\bibitem{squazzoni2020computational}
Flaminio Squazzoni, J~Gareth Polhill, Bruce Edmonds, Petra Ahrweiler, Patrycja
  Antosz, Geeske Scholz, {\'E}mile Chappin, Melania Borit, Harko Verhagen,
  Francesca Giardini, et~al.
\newblock Computational models that matter during a global pandemic outbreak: A
  call to action.
\newblock {\em Journal of Artificial Societies and Social Simulation}, 23(2),
  2020.

\bibitem{wynants2020prediction}
Laure Wynants, Ben Van~Calster, Marc~MJ Bonten, Gary~S Collins, Thomas~PA
  Debray, Maarten De~Vos, Maria~C Haller, Georg Heinze, Karel~GM Moons,
  Richard~D Riley, et~al.
\newblock Prediction models for diagnosis and prognosis of covid-19 infection:
  systematic review and critical appraisal.
\newblock {\em bmj}, 369, 2020.

\bibitem{kitchenham2007guidelines}
Barbara Kitchenham and Stuart Charters.
\newblock Guidelines for performing systematic literature reviews in software
  engineering.
\newblock 2007.

\bibitem{crooks2012introduction}
Andrew~T Crooks and Alison~J Heppenstall.
\newblock Introduction to agent-based modelling.
\newblock In {\em Agent-based models of geographical systems}, pages 85--105.
  Springer, 2012.

\bibitem{rao1995bdi}
Anand~S Rao, Michael~P Georgeff, et~al.
\newblock Bdi agents: from theory to practice.
\newblock In {\em Icmas}, volume~95, pages 312--319, 1995.

\bibitem{wolfram1983statistical}
Stephen Wolfram.
\newblock Statistical mechanics of cellular automata.
\newblock {\em Reviews of modern physics}, 55(3):601, 1983.

\bibitem{arnold2019dag}
Kellyn~F Arnold, Wendy~J Harrison, Alison~J Heppenstall, and Mark~S Gilthorpe.
\newblock Dag-informed regression modelling, agent-based modelling and
  microsimulation modelling: a critical comparison of methods for causal
  inference.
\newblock {\em International journal of epidemiology}, 48(1):243--253, 2019.

\bibitem{chao2015dynamic}
Dingding Chao, Hideki Hashimoto, and Naoki Kondo.
\newblock Dynamic impact of social stratification and social influence on
  smoking prevalence by gender: an agent-based model.
\newblock {\em Social Science \& Medicine}, 147:280--287, 2015.

\bibitem{frias2011agent}
Enrique Frias-Martinez, Graham Williamson, and Vanessa Frias-Martinez.
\newblock An agent-based model of epidemic spread using human mobility and
  social network information.
\newblock In {\em 2011 IEEE third international conference on privacy,
  security, risk and trust and 2011 IEEE third international conference on
  social computing}, pages 57--64. IEEE, 2011.

\bibitem{chatterjee2020transparency}
N~Chatterjee.
\newblock Transparency, reproducibility, and validity of covid-19 projection
  models, 2020.

\bibitem{jorgensen2018software}
Paul~C Jorgensen.
\newblock {\em Software testing: a craftsman’s approach}.
\newblock CRC press, 2018.

\bibitem{harrold2001regression}
Mary~Jean Harrold, James~A Jones, Tongyu Li, Donglin Liang, Alessandro Orso,
  Maikel Pennings, Saurabh Sinha, S~Alexander Spoon, and Ashish Gujarathi.
\newblock Regression test selection for java software.
\newblock {\em ACM Sigplan Notices}, 36(11):312--326, 2001.

\bibitem{zhu1997software}
Hong Zhu, Patrick~AV Hall, and John~HR May.
\newblock Software unit test coverage and adequacy.
\newblock {\em Acm computing surveys (csur)}, 29(4):366--427, 1997.

\bibitem{mcminn2015oracle}
E.~T. {Barr}, M.~{Harman}, P.~{McMinn}, M.~{Shahbaz}, and S.~{Yoo}.
\newblock The oracle problem in software testing: A survey.
\newblock {\em IEEE Transactions on Software Engineering}, 41(5):507--525,
  2015.

\bibitem{nguyen2008experimental}
Cu~D Nguyen, Anna Perini, and Paolo Tonella.
\newblock Experimental evaluation of ontology-based test generation for
  multi-agent systems.
\newblock In {\em International Workshop on Agent-Oriented Software
  Engineering}, pages 187--198. Springer, 2008.

\bibitem{nguyen2009testing}
Cu~D Nguyen, Anna Perini, Carole Bernon, Juan Pav{\'o}n, and John Thangarajah.
\newblock Testing in multi-agent systems.
\newblock In {\em International Workshop on Agent-Oriented Software
  Engineering}, pages 180--190. Springer, 2009.

\bibitem{padgham2005developing}
Lin Padgham and Michael Winikoff.
\newblock {\em Developing intelligent agent systems: A practical guide},
  volume~13.
\newblock John Wiley \& Sons, 2005.

\bibitem{adra2010mutation}
Salem~F Adra and Phil McMinn.
\newblock Mutation operators for agent-based models.
\newblock In {\em 2010 Third International Conference on Software Testing,
  Verification, and Validation Workshops}, pages 151--156. IEEE, 2010.

\bibitem{tiryaki2006sunit}
Ali~Murat Tiryaki, Sibel {\"O}ztuna, Oguz Dikenelli, and Riza~Cenk Erdur.
\newblock Sunit: A unit testing framework for test driven development of
  multi-agent systems.
\newblock In {\em International Workshop on Agent-Oriented Software
  Engineering}, pages 156--173. Springer, 2006.

\bibitem{zheng2005conformance}
Mao Zheng and Vangalur~S Alagar.
\newblock Conformance testing of bdi properties in agent-based software.
\newblock In {\em 12th Asia-Pacific Software Engineering Conference
  (APSEC'05)}, pages 8--pp. IEEE, 2005.

\bibitem{freedman1991testability}
Roy~S Freedman.
\newblock Testability of software components.
\newblock {\em IEEE transactions on Software Engineering}, 17(6):553--564,
  1991.

\bibitem{scholl2001agent}
Hans~Jochen Scholl.
\newblock Agent-based and system dynamics modeling: a call for cross study and
  joint research.
\newblock In {\em Proceedings of the 34th annual Hawaii international
  conference on system sciences}, pages 8--pp. IEEE, 2001.

\bibitem{nguyen2008ontology}
Cu~D Nguyen, Anna Perini, and Paolo Tonella.
\newblock Ontology-based test generation for multiagent systems.
\newblock In {\em Proceedings of the 7th international joint conference on
  Autonomous agents and multiagent systems-Volume 3}, pages 1315--1320, 2008.

\bibitem{bakar2018agent}
Najwa~Abu Bakar and Ali Selamat.
\newblock Agent systems verification: systematic literature review and mapping.
\newblock {\em Applied Intelligence}, 48(5):1251--1274, 2018.

\bibitem{blanes2009requirements}
David Blanes, Emilio Insfran, and Silvia Abrah{\~a}o.
\newblock Requirements engineering in the development of multi-agent systems: a
  systematic review.
\newblock In {\em International Conference on Intelligent Data Engineering and
  Automated Learning}, pages 510--517. Springer, 2009.

\bibitem{arora2018systematic}
Pardeep~Kumar Arora and Rajesh Bhatia.
\newblock A systematic review of agent-based test case generation for
  regression testing.
\newblock {\em Arabian Journal for Science and Engineering}, 43(2):447--470,
  2018.

\bibitem{keyes2017invited}
Katherine~M Keyes, Melissa Tracy, Stephen~J Mooney, Aaron Shev, and Magdalena
  Cerd{\'a}.
\newblock Invited commentary: agent-based models—bias in the face of
  discovery.
\newblock {\em American journal of epidemiology}, 186(2):146--148, 2017.

\bibitem{platt2020comparison}
Donovan Platt.
\newblock A comparison of economic agent-based model calibration methods.
\newblock {\em Journal of Economic Dynamics and Control}, 113:103859, 2020.

\bibitem{godefroid2012sage}
Patrice Godefroid, Michael~Y Levin, and David Molnar.
\newblock Sage: whitebox fuzzing for security testing.
\newblock {\em Queue}, 10(1):20--27, 2012.

\bibitem{fagiolo2019validation}
Giorgio Fagiolo, Mattia Guerini, Francesco Lamperti, Alessio Moneta, and Andrea
  Roventini.
\newblock Validation of agent-based models in economics and finance.
\newblock In {\em Computer Simulation Validation}, pages 763--787. Springer,
  2019.

\bibitem{thaler2019show}
Jonathan Thaler and Peer-Olaf Siebers.
\newblock Show me your properties: the potential of property-based testing in
  agent-based simulation.
\newblock In {\em Proceedings of the 2019 Summer Simulation Conference},
  page~1. Society for Computer Simulation International, 2019.

\bibitem{staats2011programs}
Matt Staats, Michael~W Whalen, and Mats~PE Heimdahl.
\newblock Programs, tests, and oracles: the foundations of testing revisited.
\newblock In {\em 2011 33rd international conference on software engineering
  (ICSE)}, pages 391--400. IEEE, 2011.

\bibitem{Petersen2015}
Kai Petersen, Sairam Vakkalanka, and Ludwik Kuzniarz.
\newblock Guidelines for conducting systematic mapping studies in software
  engineering: An update.
\newblock {\em Information and Software Technology}, 64:1--18, 2015.

\bibitem{padgham2002prometheus}
Lin Padgham and Michael Winikoff.
\newblock Prometheus: A methodology for developing intelligent agents.
\newblock In {\em International Workshop on Agent-Oriented Software
  Engineering}, pages 174--185. Springer, 2002.

\bibitem{bresciani2004tropos}
Paolo Bresciani, Anna Perini, Paolo Giorgini, Fausto Giunchiglia, and John
  Mylopoulos.
\newblock Tropos: An agent-oriented software development methodology.
\newblock {\em Autonomous Agents and Multi-Agent Systems}, 8(3):203--236, 2004.

\bibitem{bauer2001agent}
Bernhard Bauer, J{\"o}rg~P M{\"u}ller, James Odell, et~al.
\newblock Agent uml: A formalism for specifying multiagent interaction.

\bibitem{eilenberg1974automata}
Samuel Eilenberg.
\newblock {\em Automata, languages, and machines}.
\newblock Academic press, 1974.

\bibitem{laycock1993theory}
Gilbert~Thomas Laycock.
\newblock {\em The theory and practice of specification based software
  testing}.
\newblock PhD thesis, Citeseer, 1993.

\bibitem{chow1978testing}
Tsun~S. Chow.
\newblock Testing software design modeled by finite-state machines.
\newblock {\em IEEE transactions on software engineering}, (3):178--187, 1978.

\bibitem{heckel2005towards}
Reiko Heckel and Marc Lohmann.
\newblock Towards contract-based testing of web services.
\newblock {\em Electronic Notes in Theoretical Computer Science}, 116:145--156,
  2005.

\bibitem{korel1996assertion}
Bogdan Korel and Ali~M Al-Yami.
\newblock Assertion-oriented automated test data generation.
\newblock In {\em Proceedings of IEEE 18th International Conference on Software
  Engineering}, pages 71--80. IEEE, 1996.

\bibitem{clavel2002maude}
Manuel Clavel, Francisco Dur{\'a}n, Steven Eker, Patrick Lincoln, Narciso
  Mart{\i}-Oliet, Jos{\'e} Meseguer, and Jos{\'e}~F Quesada.
\newblock Maude: Specification and programming in rewriting logic.
\newblock {\em Theoretical Computer Science}, 285(2):187--243, 2002.

\bibitem{calero2006ontologies}
Coral Calero, Francisco Ruiz, and Mario Piattini.
\newblock {\em Ontologies for software engineering and software technology}.
\newblock Springer Science \& Business Media, 2006.

\bibitem{padgham2008unified}
Lin Padgham, Michael Winikoff, Scott DeLoach, and Massimo Cossentino.
\newblock A unified graphical notation for aose.
\newblock In {\em International Workshop on Agent-Oriented Software
  Engineering}, pages 116--130. Springer, 2008.

\bibitem{petri1962kommunikation}
Carl~Adam Petri.
\newblock Kommunikation mit automaten.
\newblock 1962.

\bibitem{gill1962introduction}
Arthur Gill et~al.
\newblock Introduction to the theory of finite-state machines.
\newblock 1962.

\bibitem{lee1996principles}
David Lee and Mihalis Yannakakis.
\newblock Principles and methods of testing finite state machines-a survey.
\newblock {\em Proceedings of the IEEE}, 84(8):1090--1123, 1996.

\bibitem{cheng1993automatic}
Kwang-Ting Cheng and Avinash~S Krishnakumar.
\newblock Automatic functional test generation using the extended finite state
  machine model.
\newblock In {\em 30th ACM/IEEE Design Automation Conference}, pages 86--91.
  IEEE, 1993.

\bibitem{peterson1977petri}
James~L Peterson.
\newblock Petri nets.
\newblock {\em ACM Computing Surveys (CSUR)}, 9(3):223--252, 1977.

\bibitem{shehory2001evaluation}
Onn Shehory and Arnon Sturm.
\newblock Evaluation of modeling techniques for agent-based systems.
\newblock In {\em Proceedings of the fifth international conference on
  Autonomous agents}, pages 624--631, 2001.

\bibitem{haddad1999theoretical}
Serge Haddad and Denis Poitrenaud.
\newblock Theoretical aspects of recursive petri nets.
\newblock In {\em International Conference on Application and Theory of Petri
  Nets}, pages 228--247. Springer, 1999.

\bibitem{juneidi2010survey}
Salaheddin~Juma Juneidi and GA~Vouros.
\newblock Survey and evaluation of agent-oriented software engineering main
  approaches.
\newblock {\em International Journal of Modelling and Simulation}, 30(1):1--13,
  2010.

\bibitem{peres2005experiencing}
J~Peres and Ulf Bergmann.
\newblock Experiencing auml for mas modeling: A critical view.
\newblock {\em Software Engineering for Agent-Oriented Systems, SEAS}, 2005.

\bibitem{gerber1999holonic}
Christian Gerber, J{\"o}rg Siekmann, and Gero Vierke.
\newblock Holonic multi-agent systems.
\newblock 1999.

\bibitem{smith2012ontology}
Barry Smith.
\newblock Ontology.
\newblock In {\em The furniture of the world}, pages 47--68. Brill Rodopi,
  2012.

\bibitem{moser2010ontology}
Thomas Moser, Gregor D{\"u}rr, and Stefan Biffl.
\newblock Ontology-based test case generation for simulating complex production
  automation systems.
\newblock In {\em SEKE}, pages 478--482, 2010.

\bibitem{euzenat2007ontology}
J{\'e}r{\^o}me Euzenat, Pavel Shvaiko, et~al.
\newblock {\em Ontology matching}, volume~18.
\newblock Springer, 2007.

\bibitem{hamlet2002random}
Richard Hamlet.
\newblock Random testing.
\newblock {\em Encyclopedia of software Engineering}, 2002.

\bibitem{claessen2011quickcheck}
Koen Claessen and John Hughes.
\newblock Quickcheck: a lightweight tool for random testing of haskell
  programs.
\newblock {\em Acm sigplan notices}, 46(4):53--64, 2011.

\bibitem{bailey1975mathematical}
Norman~TJ Bailey et~al.
\newblock {\em The mathematical theory of infectious diseases and its
  applications}.
\newblock Charles Griffin \& Company Ltd, 5a Crendon Street, High Wycombe,
  Bucks HP13 6LE., 1975.

\bibitem{epstein1996growing}
Joshua~M Epstein and Robert Axtell.
\newblock {\em Growing artificial societies: social science from the bottom
  up}.
\newblock Brookings Institution Press, 1996.

\bibitem{meseguer1992conditional}
Jos{\'e} Meseguer.
\newblock Conditional rewriting logic as a unified model of concurrency.
\newblock {\em Theoretical computer science}, 96(1):73--155, 1992.

\bibitem{davis1991handbook}
Lawrence Davis.
\newblock Handbook of genetic algorithms.
\newblock 1991.

\bibitem{whitley1994genetic}
Darrell Whitley.
\newblock A genetic algorithm tutorial.
\newblock {\em Statistics and computing}, 4(2):65--85, 1994.

\bibitem{deb2002fast}
Kalyanmoy Deb, Amrit Pratap, Sameer Agarwal, and TAMT Meyarivan.
\newblock A fast and elitist multiobjective genetic algorithm: Nsga-ii.
\newblock {\em IEEE transactions on evolutionary computation}, 6(2):182--197,
  2002.

\bibitem{mathur1994empirical}
Aditya~P Mathur and W~Eric Wong.
\newblock An empirical comparison of data flow and mutation-based test adequacy
  criteria.
\newblock {\em Software Testing, Verification and Reliability}, 4(1):9--31,
  1994.

\bibitem{low1999automated}
Chi~Keen Low, Tsong~Yueh Chen, and Ralph R{\'o}nnquist.
\newblock Automated test case generation for bdi agents.
\newblock {\em Autonomous Agents and Multi-Agent Systems}, 2(4):311--332, 1999.

\bibitem{cavarra2011data}
Alessandra Cavarra.
\newblock A data-flow approach to test multi-agent asms.
\newblock {\em Formal aspects of computing}, 23(1):21--41, 2011.

\bibitem{rapps1985selecting}
Sandra Rapps and Elaine~J. Weyuker.
\newblock Selecting software test data using data flow information.
\newblock {\em IEEE transactions on software engineering}, (4):367--375, 1985.

\bibitem{jia2010analysis}
Yue Jia and Mark Harman.
\newblock An analysis and survey of the development of mutation testing.
\newblock {\em IEEE transactions on software engineering}, 37(5):649--678,
  2010.

\bibitem{fink1997property}
George Fink and Matt Bishop.
\newblock Property-based testing: a new approach to testing for assurance.
\newblock {\em ACM SIGSOFT Software Engineering Notes}, 22(4):74--80, 1997.

\bibitem{goldstein1999emergence}
Jeffrey Goldstein.
\newblock Emergence as a construct: History and issues.
\newblock {\em Emergence}, 1(1):49--72, 1999.

\bibitem{miller2001acceptance}
Roy Miller and Christopher~T Collins.
\newblock Acceptance testing.
\newblock {\em Proc. XPUniverse}, 238, 2001.

\bibitem{padgham2013model}
Lin Padgham, Zhiyong Zhang, John Thangarajah, and Tim Miller.
\newblock Model-based test oracle generation for automated unit testing of
  agent systems.
\newblock {\em IEEE Transactions on Software Engineering}, 39(9):1230--1244,
  2013.

\bibitem{hadar2010empirical}
Irit Hadar, Tsvi Kuflik, Anna Perini, Iris Reinhartz-Berger, Filippo Ricca, and
  Angelo Susi.
\newblock An empirical study of requirements model understanding: Use case vs.
  tropos models.
\newblock In {\em Proceedings of the 2010 ACM Symposium on Applied Computing},
  pages 2324--2329, 2010.

\bibitem{kalaji2009generating}
Abdul~Salam Kalaji, Robert~Mark Hierons, and Stephen Swift.
\newblock Generating feasible transition paths for testing from an extended
  finite state machine (efsm).
\newblock In {\em 2009 international conference on software testing
  verification and validation}, pages 230--239. IEEE, 2009.

\bibitem{friedman2002projected}
Galit Friedman, Alan Hartman, Kenneth Nagin, and Tomer Shiran.
\newblock Projected state machine coverage for software testing.
\newblock {\em ACM SIGSOFT Software Engineering Notes}, 27(4):134--143, 2002.

\bibitem{leitner2007contract}
Andreas Leitner, Ilinca Ciupa, Manuel Oriol, Bertrand Meyer, and Arno Fiva.
\newblock Contract driven development test driven development-writing test
  cases.
\newblock In {\em Proceedings of the the 6th joint meeting of the European
  software engineering conference and the ACM SIGSOFT symposium on The
  foundations of software engineering}, pages 425--434, 2007.

\bibitem{tan2019lessons}
He~Tan, Vladimir Tarasov, and Anders Adlemo.
\newblock Lessons learned from an application of ontologies in software
  testing.
\newblock In {\em JOWO 2019, The Joint Ontology Workshops, Graz, Austria,
  September 23-25, 2019.}, volume 2518. CEUR-WS, 2019.

\bibitem{eleftherakis2011methodology}
George Eleftherakis, Petros Kefalas, and Evangelos Kehris.
\newblock A methodology for developing component-based agent systems focusing
  on component quality.
\newblock In {\em 2011 Federated Conference on Computer Science and Information
  Systems (FedCSIS)}, pages 561--568. IEEE, 2011.

\bibitem{kissoum2008recursive}
Yacine Kissoum and Zaidi Sahnoun.
\newblock A recursive colored petri nets semantics for auml as base of test
  case generation.
\newblock In {\em 2008 IEEE/ACS International Conference on Computer Systems
  and Applications}, pages 785--792. IEEE, 2008.

\bibitem{murata1989petri}
Tadao Murata.
\newblock Petri nets: Properties, analysis and applications.
\newblock {\em Proceedings of the IEEE}, 77(4):541--580, 1989.

\bibitem{dehimi2019novel}
Nour El~Houda Dehimi and Farid Mokhati.
\newblock A novel test case generation approach based on auml sequence diagram.
\newblock In {\em 2019 International Conference on Networking and Advanced
  Systems (ICNAS)}, pages 1--4. IEEE, 2019.

\bibitem{gonccalves2015mas}
Enyo Jos{\'e}~Tavares Gon{\c{c}}alves, Mariela~I Cort{\'e}s, Gustavo
  Augusto~Lima Campos, Yrleyjander~S Lopes, Emmanuel~SS Freire, Viviane~Torres
  da~Silva, Kleinner Silva~Farias de~Oliveira, and Marcos~Antonio de~Oliveira.
\newblock Mas-ml 2.0: Supporting the modelling of multi-agent systems with
  different agent architectures.
\newblock {\em Journal of Systems and Software}, 108:77--109, 2015.

\bibitem{zhang2009automated}
Zhiyong Zhang, John Thangarajah, and Lin Padgham.
\newblock Automated testing for intelligent agent systems.
\newblock In {\em International Workshop on Agent-Oriented Software
  Engineering}, pages 66--79. Springer, 2009.

\bibitem{thangarajah2011scenarios}
John Thangarajah, Gaya Jayatilleke, and Lin Padgham.
\newblock Scenarios for system requirements traceability and testing.
\newblock In {\em Autonomous Agents and MultiAgent Systems}, pages 285--292.
  IFAAMAS, 2011.

\bibitem{arcuri2011random}
Andrea Arcuri, Muhammad~Zohaib Iqbal, and Lionel Briand.
\newblock Random testing: Theoretical results and practical implications.
\newblock {\em IEEE Transactions on Software Engineering}, 38(2):258--277,
  2011.

\bibitem{shahir2012generating}
Hamed~Yaghoubi Shahir, Uwe Gl{\"a}sser, Roozbeh Farahbod, Piper Jackson, and
  Hans Wehn.
\newblock Generating test cases for marine safety and security scenarios: a
  composition framework.
\newblock {\em Security Informatics}, 1(1):4, 2012.

\bibitem{mokhati2013novel}
Farid Mokhati, Mourad Badri, and Salim Zerrougui.
\newblock A novel conformance testing technique for agent interaction
  protocols.
\newblock In {\em 2013 Science and Information Conference}, pages 485--495.
  IEEE, 2013.

\bibitem{nguyen2012evolutionary}
Cu~D Nguyen, Simon Miles, Anna Perini, Paolo Tonella, Mark Harman, and Michael
  Luck.
\newblock Evolutionary testing of autonomous software agents.
\newblock {\em Autonomous Agents and Multi-Agent Systems}, 25(2):260--283,
  2012.

\bibitem{papadakis2019mutation}
Mike Papadakis, Marinos Kintis, Jie Zhang, Yue Jia, Yves Le~Traon, and Mark
  Harman.
\newblock Mutation testing advances: an analysis and survey.
\newblock In {\em Advances in Computers}, volume 112, pages 275--378. Elsevier,
  2019.

\bibitem{just2014defects4j}
Ren{\'e} Just, Darioush Jalali, and Michael~D Ernst.
\newblock Defects4j: A database of existing faults to enable controlled testing
  studies for java programs.
\newblock In {\em Proceedings of the 2014 International Symposium on Software
  Testing and Analysis}, pages 437--440, 2014.

\bibitem{zhou2016map}
Xin Zhou, Yuqin Jin, He~Zhang, Shanshan Li, and Xin Huang.
\newblock A map of threats to validity of systematic literature reviews in
  software engineering.
\newblock In {\em 2016 23rd Asia-Pacific Software Engineering Conference
  (APSEC)}, pages 153--160. IEEE, 2016.

\end{thebibliography}

\end{document}